\documentclass[%
 aip,
 amsmath,amssymb,
preprint,%
]{revtex4-1}

\newcommand{\nid}{\noindent}
\usepackage{bm}
\usepackage{acronym}
\acrodef{VSFG}[VSFG]{Vibrational sum frequency generation }
\acrodef{SI}[SI]{Supporting Information}
\acrodef{ssVVCF}[ssVVCF]{surface-specific velocity-velocity correlation function formalism}
\acrodef{RMSE}[RMSE]{root  mean  square  error}
\acrodef{DFT}[DFT]{density functional theory}
\acrodef{SR}[SR]{short range}
\acrodef{LR}[LR]{long range}
\acrodef{HB}[HB]{hydrogen bonded}
\acrodef{ML}[ML]{machine-learning} 
\acrodef{MD}[MD]{molecular dynamics}
\acrodef{XC}[XC]{exchange correlation}
\acrodef{CMD}[CMD]{centroid molecular dynamics} 
\acrodef{PES}[PES]{potential energy surface}
\acrodef{NQE}[NQE]{nuclear quantum effects}
\acrodef{HDNNP}[HDNNP]{high-dimensional neural networks}
\acrodef{PICG}[PICG]{path integral coarse-grained}
\acrodef{SA-GPR}[SA-GPR]{symmetry-adapted Gaussian process regression}
\acrodef{TRPMD}[TRPMD]{ thermostated ring polymer molecular dynamics}
\usepackage{graphicx}

\begin{document}

\preprint{AIP/123-QED}

\title{Fully First-Principles Surface Spectroscopy with Machine Learning}

\author{Yair Litman}%
\affiliation{Yusuf Hamied Department of Chemistry,  University of Cambridge,  Lensfield Road,  Cambridge,  CB2 1EW,UK}
\affiliation{Max Planck Institute for Polymer Research, Ackermannweg 10, 55128 Mainz, Germany}
\email{yl899@cam.ac.uk}

\author{Jinggang Lan}
\affiliation{Department of Chemistry, New York University, New York, NY, 10003, USA}
\affiliation{Simons Center for Computational Physical Chemistry at New York University, New York, NY, 10003, USA}
\author{Yuki Nagata}
\affiliation{Max Planck Institute for Polymer Research, Ackermannweg 10, 55128 Mainz, Germany}
\author{David M. Wilkins}
\affiliation{Centre for Quantum Materials and Technology, School of Mathematics and Physics, Queen’s University Belfast, Belfast BT7 1NN, Northern Ireland, United Kingdom}
\email{d.wilkins@qub.ac.uk}

\begin{abstract}
Our current understanding of the structure and dynamics of aqueous interfaces at the molecular level has grown substantially in the last few decades due to the
continuous development  of surface-specific spectroscopies,
such as vibrational sum-frequency generation (VSFG).
Similarly to what happens in other spectroscopies, to extract all of the information encoded in the VSFG spectra we must turn to atomistic simulations.
The latter are conventionally based either on empirical force field models,
which cannot describe bond breaking and formation or systems with a complex electronic structure, or on \textit{ab initio} calculations which are difficult to statistically converge due to their computational cost. These limitations ultimately hamper our understanding of aqueous interfaces. In this work, we overcome these constraints by combining two machine learning techniques, namely high-dimensional neural network interatomic potentials and symmetry-adapted Gaussian process regression, to simulate the SFG spectra of the water/air interface with \textit{ab initio} accuracy.
Leveraging a data-driven local decomposition of atomic environments, we develop a simple scheme that allows us to obtain VSFG spectra in agreement with current experiments. Moreover, we identify the main sources of inaccuracy and establish a clear pathway towards the modelling of surface-sensitive spectroscopy of complex interfaces.
\end{abstract}

\maketitle

Interfaces of aqueous solutions are ubiquitous in nature and play a crucial role in many  important processes, such as (electro)catalytic applications   \cite{Magnussen_Jacs_2019}, atmospheric aerosol–gas exchanges \cite{Jubb_AnnRev_2012}, and mineral dissolution \cite{Schott_RevMinGeo_2009}. In particular, the water/air interface has received enormous attention from the scientific community in the last few decades since it represents arguably the most simple and important interface with a hydrophobic surface and serves as a baseline from which more complex aqueous interfaces can be analyzed, interpreted and rationalized \cite{Gonella_NatRev_2021}. Water presents many interesting properties, such as anomalously high surface tension and a non-monotonic temperature dependence of its density with a maximum at 4$^\circ$C \cite{Petterson_ChemRev_2016}. These unique properties are attributed to its unusually strong hydrogen-bond (HB) networks. A fundamental understanding of the properties and reactivity of aqueous interfaces thus demands a molecular-level description that can capture the fine energetic balances governing the structural and dynamical characteristics of the HB networks \cite{Fecko_Science_2003,Bjorneholm_ChemRev_2016}.

Techniques that isolate the signal from the relatively few surface molecules at the surfaces from the enormous contribution due to the bulk are essential to study interfaces. \ac{VSFG}  belongs to a selected class of spectroscopic techniques with the capability of probing such interfaces with molecular-level and chemical sensitivity \cite{Shen1989}. In \ac{VSFG} experiments, IR and UV-visible pulses are spatially and temporally overlapped and a signal generated by the sample 
at the sum of the frequencies of the incoming radiations is measured. The signal is determined by the second order susceptibility 
 of the sample, $\chi^{(2)}$. Since $\chi^{(2)}$ is zero in centrosymmetric (bulk) environments, it naturally provides information that emerges exclusively from interfacial molecules. 
The \ac{VSFG} spectra of the O-H stretch mode at aqueous interfaces presents broad and featureless bands. Thus,  experimental data alone is insufficient to unambiguously disentangle spectral components, and atomistic simulations are required for a microscopic understanding.
Atomistic simulations have played an important role in settling the long-standing debate regarding the absence of a positive signal below 3200 cm$^{-1}$ for H$_2$O \cite{Nihonyanagi_JCP_2015,Tang_ChemRev_2020}, in determining the 
dielectric function profile across the water interface \cite{Chiang_PNAS_2022}, and in elucidating the pH-dependent structure at the fluorite/water interface 
\cite{Khatib_SciRep_2016}.

 The theoretical calculation of \ac{VSFG} spectra is more challenging than that of more traditional spectroscopies such as linear IR and Raman, since relatively long simulation times, on the order of nanoseconds, are required to converge the statistics so that the signal in the bulk-like (centrosymmetric) regions vanishes \cite{Morita_JPCB_2002,Moberg_JPCB_2018}. 
 In this context, several approximations have been reported with different degrees of success when dealing with aqueous solutions.  A first group of studies is based on approximations of the \ac{PES} by classically polarizable force fields \cite{Nagata_JPCL_2013,Moberg_JPCB_2018,Morita_JPCB_2002,Khatib_JPCC_2016}. Many important insights have been obtained with these approaches such as the determination of molecular orientation at interfaces  \cite{Yu_PhysRevLett_2022} or the 
 proof that chiral \ac{VSFG} exclusively probes the first hydration shell around biomolecules   \cite{Konstantinovsky_ACS_2022}. 
 However, these calculations 
 cannot describe the rearrangement of chemical bonds nor systems with complex electronic structure.
 A second important group is based on the combination of \textit{ab initio} potentials at the \ac{DFT} level with approximations of the dipole and polarizability surfaces by simple models such as the \ac{ssVVCF}  \cite{ssVVCF,ssVVCF_Sulpizi} or frequency maps  \cite{Ni_JCP_2015}. These approximations allow converged results to be obtained with relatively shorter simulations and deliver overall appropriate spectral shapes, including the absence of a positive signal below 3200 cm$^{-1}$ for H$_{2}$O \cite{Nihonyanagi_JCP_2015,ssVVCF}. However, the oversimplified description of dipole and polarizability surfaces cannot distinguish between different polarization combinations, which is required for estimating the molecular orientation and studying the depth profile of molecules
 \cite{ChunChieh_JPCB_2022}. They are also unable to capture important spectral features originating from vibrational coupling 
\cite{Tian_JACS_2009,Nihonyanagi_JACS_2011,Yamaguchi_JCPL_2022,Chiang_chemArxiv_2023}.
Going beyond these approximations and obtaining a \textit{fully} \textit{{ab initio}} MD \ac{VSFG} spectra of liquids  is a long-sought goal in the SFG community, but due to prohibitively large computational expense, statistically converged simulations remain elusive \cite{Ojha_Molecules_2020}.

The vibrational resonant component of the second order susceptibility, $\chi^{(2),R}_{pqr}$, in a electronically non-resonant condition, can be computed as \cite{Morita_JPCB_2002}

\begin{equation}\label{eq:TCF}
\begin{split}
\chi^{(2),R}_{pqr} (\omega_\text{IR}) =i\int_0^\infty dt e^{-i\omega_\text{IR}t}\langle \alpha_{pq}(t)P_r(0)\rangle,
 \end{split}
\end{equation}
\nid 
where $\alpha_{pq}$ is the $pq$ component of the polarizability  tensor, $\omega_\text{IR}$ is frequency of the IR pulse, and $P_{r}$ is the $r$ component of the polarization vector. Hereafter, we refer to the resonant part exclusively so the $R$ supraindex is omitted. The evaluation of Eq. \ref{eq:TCF} requires an accurate representation of three objects, namely, i) \ac{PES}, ii) polarization surface ($P$-S) and iii) polarizability surface ($\alpha$-S). 
Moreover, the light mass of hydrogen atoms, abundant in aqueous solutions, demands a quantum mechanical treatment of the molecular motions \cite{
Markland_NatRevChem_2018}. In this work, the \ac{PES} is evaluated using either \ac{DFT} or Behler-Parrinello \ac{HDNNP} \cite{HDNNP}, $P$-S and $\alpha$-S are evaluated by a \ac{SA-GPR} scheme that enables the prediction of  tensorial  quantities  of  arbitrary  order \cite{SA-GPR,Deringer2021}, and the time-evolution of the nuclei using classical
\ac{MD}, and \ac{CMD}~\cite{CMD_I,CMD_II} in its partially adiabatic approximation \cite{Hone_JCP_2006}. The latter method is able to capture the most important \ac{NQE} in condensed phases, such as zero-point-energy and tunneling, and is suitable for spectroscopic simulations of aqueous solutions at room temperature~\cite{Markland_NatRevChem_2018, Marsalek_JCPL_2017}.

The modelling of aqueous solution/air interfaces is normally performed by simulating the system under study in a slab geometry, in which the system extends infinitely along two dimensions and has finite size along the third dimension where it is sandwiched by regions of vacuum. This geometry gives two interfaces that will generate \ac{VSFG} signals with opposite signs that cancel each other and lead to a vanishing signal. Thus, it is not possible to obtain any meaningful spectra if one uses $P$ and $\alpha$ of the total slab in Eq. \ref{eq:TCF}.
As we will discuss below, our \ac{ML} approach offers an elegant and data-driven solution to this issue.

\subsection*{Training and Validation of ML models}

SA-GPR predictions of a selected property, $\bm{y}$, can be expressed as a linear combination,
\begin{equation}\label{eq:SA-GPR}
\begin{split}
\bm{y}(\mathcal{X})=
\sum_I^{N_t}
\sum_{i,j}^{N} w^I k^{(\lambda)}(\mathcal{X}_i,\mathcal{X}_j^I),
 \end{split}
\end{equation}
\nid where  $k^{(\lambda)}(\mathcal{X}_i,\mathcal{X}^I_j)$ is the 
kernel function that evaluates the similarity
between the $i^{\rm th}$ atomic environment of the trial
configuration, and the $j^{\rm th}$ atomic environment of the $I^{\rm th}$ reference configuration. $N_t$ and $N$ represent the number of training points and number of atomic environments, respectively, and  
$\{w^i\}$ are the weights determined during a training procedure \cite{StatisLearning}.
$\lambda$ identifies an orthogonal subspace of dimension
2$\lambda$ + 1, according to SO(3) algebra \cite{Weinert_1980}, such that the modeling of  $P$ ($\alpha$) requires the $\lambda=1$ ($\lambda=0,2$) components.
The training of the  $P$-S and $\alpha$-S  was performed using  \ac{DFT} at the PBE and PBE0 level, and POLY2VS, a polarizable water force field developed by Tanimura and co-workers~\cite{POLY2VS}. The PBE and PBE0 data sets are obtained from first principles and constitutes the ultimate target of this work. The POLY2VS data set provides access to a molecular decomposition of the $\alpha$ and $P$ quantities allowing us to critically assess the performance of the \ac{ML} models. Moreover, we considered two types of data sets, one made up exclusively of bulk structures and a second one made of water clusters from monomers up to hexamers (see a more detailed description of the data set and the training procedure in the  Methods section below). 
We used the water \ac{HDNNP} generated by Schran and co-workers using an active learning strategy \cite{Schran_JCP_2020}. This model was trained on 814 reference data points including liquid, ice and slab structures at the revPBE0-D3 theory level,  showing excellent agreement with the reference data. This level of theory is known to describe the structure and dynamics of liquid water accurately \cite{Marsalek_JCPL_2017,Ohto_JPCL_2019}.  To better evaluate  the loss in accuracy in going from the \textit{ab initio} \ac{PES} to the \ac{HDNNP}, we also considered the
\ac{HDNNP} published by some of the authors trained on slab structures at the revPBE-D3 theory level \cite{Litman_arxiv_2022}.

In table \ref{tab:datasets}, we summarize the different models considered in this work. To enable comparisons between data sets, the error estimates are computed as the \ac{RMSE} percentage of the intrinsic deviation of the data set and expressed per water molecule. 
The \ac{SA-GPR} models can accurately learn $P$ and $\alpha$ with errors below a few percentages of the intrinsic variation in the training set.
These values represent an error of $5.4\times 10^{-4}$ D/atom ($7.6\times 10^{-4}$ D/atom) and $7.8\times 10^{-4}$ $\AA^3$/atom ($1.1\times 10^{-3}$ $\AA^3$/atom) for  $|P|$ and Tr[$\alpha$], respectively, for the ML-POLY-B (ML-POLY-A) model.
In the \ac{SI}, we report the correlation plots and learning curves of the ML models, together with a brief analysis of the training errors.

\begin{table}[tbhp]
\centering
\caption{Root mean squared error (RMSE) for SA-GPR models of polarization per atom (in Debye) and polarizability per atom (in $\text{\AA}^{3}$). Numbers in brackets give the RMSE as a percentage of the intrinsic deviation in the training set. }\label{tab:datasets}
\begin{tabular}{c|c|c|c}
Model Name & Configurations & Polarization ($\boldsymbol{P}$)  & Polarizability ($\boldsymbol{\alpha}$) \\
\hline
 ML-POLY-A & Bulk Water & $7.6\times 10^{-4} ~(1.7\%)$ & $2.7\times 10^{-3} ~(13\%)$ \\
 ML-POLY-B & Water clusters &  $5.4\times 10^{-4}~(0.5\%)$  & $2.2\times 10^{-3}~(3\%)$ \\
 ML-PBE-A &  Bulk Water & $1.0\times 10^{-3} ~(0.8\%)$ & $2.6\times 10^{-3}~(13\%)$ \\
 ML-PBE0-A &  Bulk Water & $2.2\times 10^{-3} ~(0.8\%)$ & -- \\%
\hline
\end{tabular}

\end{table}

\subsection*{ML aided simulation of \ac{VSFG} spectra}

 As previously noted,  $\chi^{(2)}$ for a slab geometry is zero due to a cancellation between the contributions of opposite interfaces. Assuming that the slab thickness is large enough to accommodate a bulk region in the middle and that a molecular decomposition of $P$ is available, it is standard practice to set to zero or flip the sign of the molecular dipoles below the centre of mass of the slab to avoid this cancellation \cite{ssVVCF,ssVVCF_Sulpizi}.
However,  molecular dipoles are not observables and therefore they are arbitrary in nature.  In \textit{ab initio} calculations the 
molecular decomposition can be achieved by using maximally localized Wannier functions \cite{Wan_PRL_2015,Ojha_Molecules_2020}, which incurs the difficulty of requiring direct  \textit{ab initio} simulations~\cite{marx_book}.
Here instead, we utilize an unbiased data-driven approach.
Since we are using an atom-centred decomposition to represent the atomic environments, by rearranging the sums in Eq. \ref{eq:SA-GPR}, SA-GPR predictions 
can be expressed as a sum of molecular predictions as 
\begin{equation}\label{eq:GPR1}
\begin{split}
\bm{y}(\mathcal{X})=\sum_{\gamma}\bm{y}(\mathcal{X}_\gamma),
 \end{split}
\end{equation}
\nid where $\gamma$ represents the set of indices that corresponds to a given molecule. Thus, SA-GPR can be applied to the calculation of SA-GPR using Eq. \ref{eq:TCF} for slab geometries by applying the following modification to Eq. \ref{eq:GPR1} for the $P$ predictions

\begin{equation}\label{eq:filter1}
\begin{split}
P(\mathcal{X})=\sum_{\gamma}
g(z_\gamma)P(\mathcal{X}_\gamma),
 \end{split}
\end{equation}
\nid where
the surface plane is assumed to be parallel to the $xy$ plane, $z_\gamma$ is $z$ coordinate of the $\gamma$ molecule, and 
\begin{equation}\label{eq:filter2}
\begin{split}
g(z,z_1,z_2) &= \text{sgn}(z)\\& \times
     \begin{cases}
       1 & |z-z_\text{cm}|\ge  z_1\\
       \cos(\frac{(z_1-|z-z_\text{cm}|) \pi}{2(z_1-z_2)}) & z_1\ge|z-z_\text{cm}|\ge  z_2\\
       0 & z_2\ge|z-z_\text{cm}|
     \end{cases},
 \end{split}
\end{equation}
\nid where $z_\text{cm}$ is the z-coordinate of the slab center of mass, and $z_1$ and $z_2$ are two parameters that define the transition between the interfacial and bulk regions.

\begin{figure*}%
\centering
\includegraphics[width=1.0\linewidth]{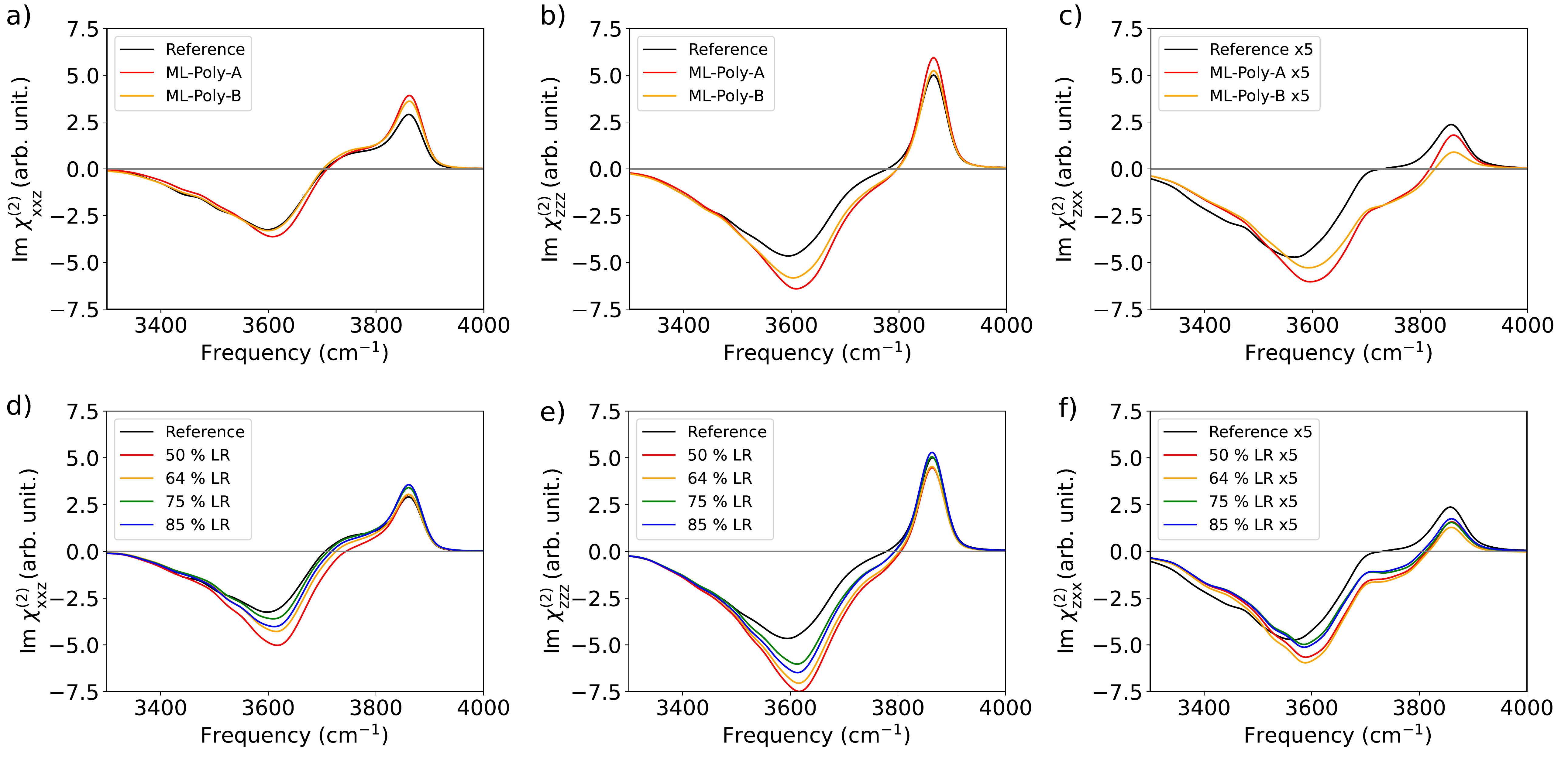}
\caption{a-c) Imaginary part of
non-zero and independent $\chi^{(2)}$ components of the water/air interface using the POLY2VS model (reference) and ML-POLY-A and ML-POLY-B \ac{SA-GPR} models. d-f) 
Same as a-c) but using ML-POLY-A augmented by different amounts of \ac{LR} contributions. $\chi^{(2)}_
\text{zxx}$ spectra are multiplied by a factor of 5 to ease visualization.
}
\label{fig:GPRonPOLY2vs}
\end{figure*}

Global accuracy estimators, such as the \ac{RMSE} and absolute error presented above, are 
useful indicators of the overall performance of \ac{ML} models.
However, when the learned quantities are used as a proxy for further calculations, as we do here for the calculation of $\chi^{(2),R}$,
it is normally not possible to translate RMSE values directly into an objective accuracy measure of the final target. This is mainly because of the difficulty in performing a rigorous error propagation of the uncertainties in Eq. \ref{eq:TCF} but also due to the lack of a unique set of descriptors to quantify how ``good'' a predicted spectrum is.  In this work, we take what we consider to be the strictest validation test and  in Fig. \ref{fig:GPRonPOLY2vs} compare the \ac{VSFG} spectra of water obtained with our \ac{ML} models against the corresponding reference  POLY2VS spectra. We note that this can only be done because POLY2VS by construction has a molecular decomposition
of $P$ and $\alpha$ that enables the application of the filter function (Eq. \ref{eq:filter2}). The Im$\chi^{(2)}$ spectrum of the water/air interface presents, irrespective of the polarization, a sharp positive peak centered at 3700 cm$^{-1}$ and a broad band in the 3200-3500 cm$^{-1}$ region with negative amplitude signal. The peak is due to the dangling (free) O-H group of the topmost interfacial water layer, 
and the band is associated with O-H groups of
the interfacial water which form hydrogen-bonds with other water molecules. The sign of the amplitudes in the Im$\chi^{(2)}$ spectrum can be directly associated with the orientation of the water molecules. More specifically, 
positive (negative) amplitudes imply that the
O-H group points  towards the air  (towards the bulk) phase.  $\chi^{(2)}_\text{pqr}$ presents only 3 independent elements:
$\chi^{(2)}_\text{zzz}$,
$\chi^{(2)}_\text{zxx}$, and
$\chi^{(2)}_\text{xxz}$.
$\chi^{(2)}_\text{zxx} =\chi^{(2)}_\text{zyy}=\chi^{(2)}_\text{xzx}=\chi^{(2)}_\text{yzy}$, and $\chi^{(2)}_\text{xxz}=\chi^{(2)}_\text{yyz}$  due to the $xy$ plane being isotropic and the symmetry of the polarizability tensor. The rest of the elements are exactly zero except for large chiral environments \cite{Jie_PNAS_2005,Perets_Langmuir_2022}. In the case of the water/air interface, the non-zero Im$\chi^{(2)}$ tensor components differ essentially by the total intensity, the relative intensity of the free O-H and hydrogen-bonded O-H regions, and the presence or absence of a shoulder on the lower frequency side of the free O-H peak. The components have also a markedly different temperature dependence \cite{Moberg_JPCB_2018}.

We proceed with a systematic evaluation of the \ac{ML} predictions of the \ac{VSFG} spectra. First, we consider one \ac{PES} given by the POLY2VS force-field and focus on the impact of different training sets used to create the \ac{SA-GPR} models. Fig. \ref{fig:GPRonPOLY2vs}a, Fig \ref{fig:GPRonPOLY2vs}b, and Fig \ref{fig:GPRonPOLY2vs}c show $\chi^{(2)}_\text{xxz}$, $\chi^{(2)}_\text{zzz}$, and $\chi^{(2)}_\text{zxx}$, respectively. 
The ML-POLY-A and ML-POLY-B models have similar accuracy in their predictions of $\chi^{(2)}_\text{xxz}$ and $\chi^{(2)}_\text{zzz}$ and provide a semi-quantitative agreement with the reference spectra. This a rather surprising result since the ML-POLY-A model was trained in bulk water conformations and therefore has no information in the training set on interfacial water structures, whereas the ML-POLY-B model was trained on water cluster conformations, and had only limited information on fully solvated water molecules.  The prediction of $\chi^{(2)}_\text{zxx}$ is considerably worse than the other two tensor components with ML-POLY-A performing only slightly better than ML-POLY-B for the free O-H peak. By performing cross-predictions using $P$/$\alpha$ from the ML model and $\alpha$/$P$ from the reference, we identified the $\alpha$ predictions as the source of the observed inaccuracy (see Fig. S12 in the \ac{SI}). %
This result highlights the limitation of using RMSE values and correlations plots as error estimators as all components of $\alpha$ show a comparable accuracy.

Electrostatic  \ac{LR} effects are known to play an important role at interfaces due to unbalanced  interactions that can build up in the presence of a broken translation symmetry~\cite{Niblett_JCP_2021}.
The implications of using short-range models on structural properties of the water/air interface,  such as 
 orientations of water molecules or average density   are well documented \cite{Yue_JCP_2021,Gao_NatCom_2022}. \ac{ML} models that include \ac{LR} effects are also known to be more accurate \cite{Niblett_JCP_2021,Grisafi_JCP_2019}. However, the possible impact on non-linear spectroscopic responses such as $\chi^{(2)}$ is less clear.
 To analyze this effect, we trained a long-distance equivariant (LODE) model based on the local value of an atom-density potentials~\cite{Grisafi_JCP_2019}.
 In Fig. \ref{fig:GPRonPOLY2vs}c-d, we show the predictions using a combination of \ac{SR} and \ac{LR}, since the combined model normally delivers a superior accuracy than the individual ones. In particular, we considered different amounts of \ac{LR} contributions in the range 50\%-85\%. The best combination across all the tensor components is obtained for a 25/75 \ac{SR}/\ac{LR} model, but the performance is very similar to that of \ac{SR} models. These results unequivocally demonstrate that \ac{LR} effects on the description of $P$ and $\alpha$ do not have a huge impact on the $\chi^{(2)}$ spectra.

\subsection*{First principles simulation of \ac{VSFG} spectra}

\begin{figure*}[tbhp]
\centering
\includegraphics[width=1.0\linewidth]{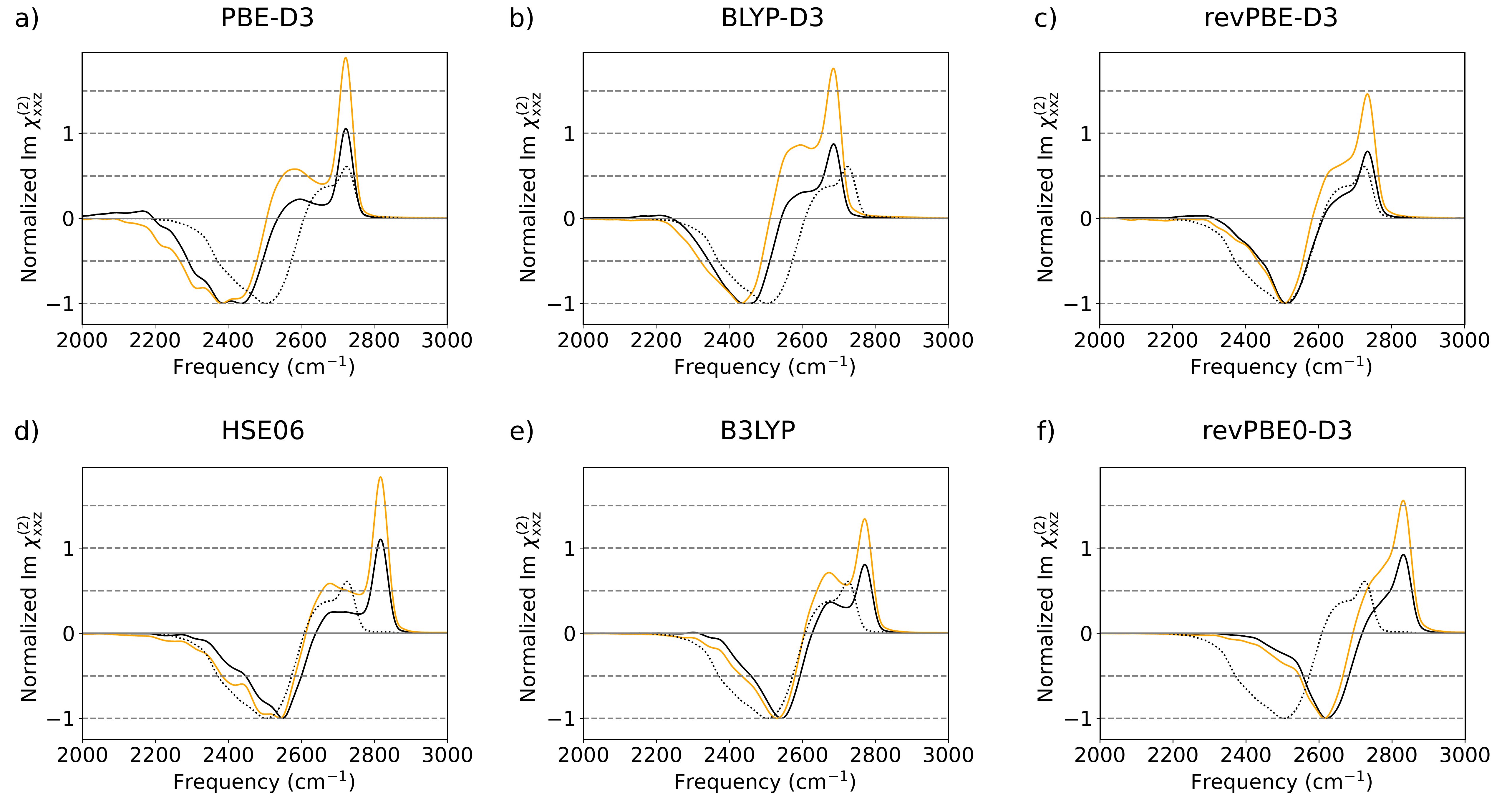}
\caption{Simulated SFG spectra, $(\chi^{(2)}_{xxz})$, of the water/air interface  using trajectories obtained with \textit{ab initio} \ac{PES} and
ML-PBE-A and ML-POLY-A are depicted by orange and black lines, respectively. Experimental spectra are corrected with Fresnel factors, assuming the Lorentz model  for the interfacial dielectric constant is depicted in black dotted lines~\cite{Xiaoqing_JCP_2023}. }
\label{fig:GPRonabinitio}
\end{figure*}

Having established the suitability of the training set and the methodology,
we now consider the spectral changes induced by different \textit{ab initio} \ac{PES} for fixed $P$ and $\alpha$ surfaces.
In Fig. \ref{fig:GPRonabinitio}, we present the computed \ac{VSFG} spectra using ML-PBE-A and ML-POLY-A response surfaces and six different \textit{ab initio} potential energy surfaces, namely PBE-D3, BLYP-D3, revPBE-D3, HSE06, B3LYP and revPBE0-D3. The trajectories were available from a previous work \cite{Ohto_JPCL_2019}. Due to the limited length of the simulations, it was necessary to neglect intermolecular terms in Eq. \ref{eq:TCF} to obtain a reasonable converged spectrum (see more details in the SI). Thus, the width of the negative band is poorly described due to the absence of intermolecular couplings~\cite{ssVVCF}. In all cases except for PBE-D3, the spectra do not show a positive signal below 2200 cm$^{-1}$ in agreement with the latest measurements and simulations since 2015 \cite{Nihonyanagi_JCP_2015,Tang_ChemRev_2020}. In the case of D2O, the experimental free O-D amplitude is approximately 50\% smaller than of the \ac{HB} O-D band. For all the considered \ac{PES}s, the ML-PBE-A model overestimates (underestimates) the intensity of free O-H peak (\ac{HB} O-H band), while the ML-POLY-A model predicts relative intensities in better agreement with the experiments. The better performance of ML-POLY-A is a direct consequence of the fact that the reference POLY2VS dipole and polarizability surfaces were fitted to reproduce CCSD/aug-cc-pVQZ, rather than \ac{DFT}, reference values. 
 The calculations with GGA \ac{XC} functionals   (PBE-D3, BLYP and revPBE-D3) are artificially red-shifted in comparison to the results obtained with hybrid ones 
(HSE06, B3LYP and revPBE0-D3) in agreement with previous approximations based on  the \ac{ssVVCF} methodology \cite{Ohto_JPCL_2019}, and results on bulk water \cite{Gillan_JCP_2016,Marsalek_JCPL_2017}. 
    Inclusion of \ac{NQE}s are known to induce a frequency red-shift when compared  to the  corresponding classical nuclei counterpart spectra \cite{Markland_NatRevChem_2018,Raz_FarDis_2020}. Thus, the frequency agreement between the revPBE-D3 \ac{PES} and the experimental data is a fortuitous error compensation \cite{Marsalek_JCPL_2017}.
    
    The intramolecular vibrational coupling of water molecules in which one O-H is a hydrogen-bond donor and the other one is free has a distinctive spectral feature associated with a shoulder on the free O-H peak. This shoulder has been assigned to the asymmetric stretching mode of interfacial water molecules \cite{Schaefer_JPCL_2016,Stiopkin_Nature_2011} and contributions to this shoulder from water molecules forming two hydrogen bonds have been shown to be minor due to cancellation of inter- and intramolecular contributions \cite{Moberg_JPCB_2018}.
PBE-D3, BLYP-D3, HSE06 and B3LYP spectra present a shoulder so separated from the free O-H peak that it can be regarded as a separate peak. On the contrary, this spectral feature is correctly displayed by the  revPBE0-D3 and revPBE-D3 spectra. 
Moreover, the differences in performance for  revPBE0-D3 and revPBE-D3 are relatively small and mainly impact the intensity of the free O-D peak and an overall frequency shift. These results show that even hybrid functionals considerably overestimate the H-bond strength, unlike in bulk water~\cite{Guidon_JCP_2008}.

\begin{figure*}%
\centering
\includegraphics[width=.95\linewidth]{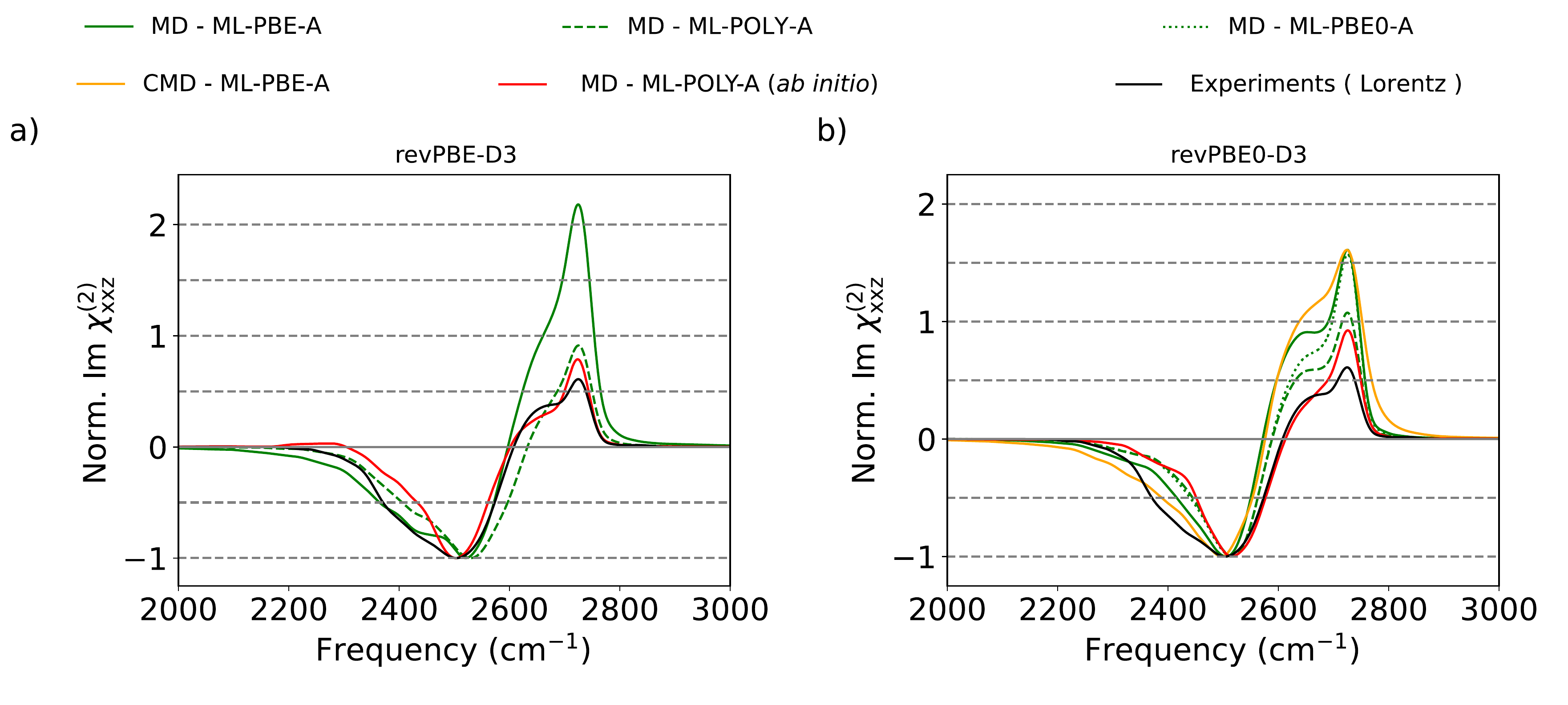}
\caption{Normalized $\chi^{(2)}_{xxz}$  spectra of the D$_2$O/air interface at 300K.   
Simulated spectra using revPBE-D3 (left) and revPBE0-D3 (right) \ac{XC} functionals.
Classical \ac{MD} simulations using \ac{HDNNP}s are presented  
by solid (ML-PBE-A), dashed (ML-POLY-A), and dotted (ML-PBE0-A) green lines.
\ac{CMD} simulations are depicted with solid orange lines while
results based direct \textit{ab initio} PES are depicted with red lines.
Experimental spectra are depicted with solid black lines and horizontal gray lines have been added to guide the eye. 
Spectra have been rigidly shifted to match experimental spectra. The values of the frequency shifts for 
revPBE0-D3 (revPBE-D3) are:
MD-\ac{HDNNP}       -138  cm$^{-1}$ ( +4 cm$^{-1}$),
MD-\textit{ab initio}   -105  cm$^{-1}$ ( -10 cm$^{-1}$), and
\ac{CMD}-\ac{HDNNP} -57   cm$^{-1}$.
 Experimental spectra are corrected by the appropriate Fresnel factors, assuming the Lorentz model (black dotted lines) for the interfacial dielectric constant \cite{Xiaoqing_JCP_2023}. To allow a visual comparative analysis,  the spectra are normalized such that the \ac{HB} O-D band has an intensity of unity. 
 }
\label{fig:experiments}
\end{figure*}

Finally, we show fully-\ac{ML} predictions for the Im$\chi^{(2)}_\text{xxz}$ spectra for D$_2$O/air at 300K in Fig. \ref{fig:experiments} (Re$\chi^{(2)}_\text{xxz}$ is reported in the SI). We focus on revPBE0-D3 and revPBE-D3 \ac{XC} functionals since they show the best agreement with experiments. 
The solid red and  dashed green curves show the predictions using the ML-POLY-A model  and \textit{ab intio} and \ac{HDNNP} \ac{PES} surfaces, respectively. 
The overall agreement between the spectra for each \ac{XC} functional is remarkable when considering the computational costs associated with each type of calculation. The revPBE0-D3 \ac{HDNNP} result shows a free O-D peak with a more pronounced free O-D peak shoulder and a blue-shift of 33 cm$^{-1}$ with respect to the \textit{ab initio} one.
Conversely, the revPBE0-D3 \ac{HDNNP} spectra show a less pronounced free O-D peak shoulder and and a red-shift of -14 cm$^{-1}$. Both \ac{HDNNP}  were trained without an explicit treatment of \ac{LR} interactions which are known, as mentioned previously, to be responsible for the net orientation of the water molecules in the bulk region. We believe that these artefacts are responsible for the discrepancies with fully \textit{ab initio} spectra. 
In particular, the spurious additional orientation of water molecules, which happens in opposite directions for 
revPBE-D3 and revPBE0-D3 \ac{HDNNP}s,
as shown in Fig. S14 in the \ac{SI}, might be responsible for increasing and decreasing the splitting between free and \ac{HB} O-D vibrations of water molecules at the topmost layer. For all tested cases, the theoretical predictions overestimate the intensity ratio between the free O-D and the hydrogen-bonded band.  Such underestimation has been reported by Paesani and co-workers to be expected when neglecting induction effects arising from interactions between individual molecules \cite{Medders_JACS_2016}. Since SA-GPR can capture local induction effects, the large discrepancy is attributed to the quality of the underlying reference data. Once more, the ML-POLY-A model outperforms the ML-PBE-A. For the revPBE0-D3 case, we also tested the the ML-PBE0-A, which uses PBE0 as reference data for $P$. While it represents an improvement with respect to  revPBE-D3, particularly on the description of free O-D shoulder, it still underperforms the ML-POLY-A results.

So far we have analyzed simulations where the nuclei were assumed to behave classically, and the spectra have been rigidly red-shifted to account for the ignored \ac{NQE}s. On the right panel of Fig. \ref{fig:experiments}, we show the results obtained with \ac{CMD}, which is a well-established method to simulate the vibrational spectra of condensed phase systems at room temperature including an approximated description of \ac{NQE}s~\cite{Habershon_JCP_2008,Rossi_JCP_2014}.
\ac{CMD} is based on the path integral formulation of quantum mechanics \cite{Feynman} and in this method the nuclei are evolved according to classical equations of motion on the so-called centroid potential of mean force.  \ac{NQE}s induce a small broadening of free O-D peak and  \ac{HB}-band and a red-shift of 81 $^{-1}$ with respect to the corresponding classical nuclei simulations. However, the \ac{CMD} spectrum is still 57 cm$^{-1}$ blue-shifted with respect to the experimental result.
We also obtained the \ac{VSFG} spectra using the \ac{TRPMD}\cite{TRPMD} method which is known to deliver more accurate frequencies than \ac{CMD} at 300 K since it 
does not suffer from the curvature problem~\cite{Ivanov_JCP_2010,Althorpe_review_2021,Trenins_JCP_2019,Rossi_JCP_2014} (see Fig. S16 in the SI).
By comparing the \ac{CMD} and \ac{TRPMD} results, we deduce
that the curvature problem is responsible for an additional 10 cm$^{-1}$ red-shift. By accounting for the  33 cm$^{-1}$ blue-shift induced by the lack of \ac{LR} effects in the \ac{HDNNP} discussed previously,  we conclude  that the error introduced by DFT in the revPBE0-D3 \ac{XC} functional approximation is 
within the theoretical limit of quantum-statistic classical-dynamics methods which overestimate high-frequency modes by about 50 cm$^{-1}$ \cite{Althorpe_review_2021,Trenins_JCP_2019}.

\subsection*{Discussion}

The combination of two different types of \ac{ML} algorithms, namely \ac{HDNNP}s and 
\ac{SA-GPR}, has allowed us to describe the water/air interface and its second-order response from first principles at an affordable computational cost. 
The training of the 
\ac{ML} models were performed in a general fashion,  without employing specific information of the system nature, such as physical constraints designed for water, mapping models for water \cite{Ni_JCP_2015}, or the use of a $\Delta$\ac{ML} procedure based on an available surrogate model \cite{Qu_DigDis_2022}. While at the moment force-fields tailored to describe water outperform the presented results \cite{Moberg_JPCB_2018,Xiaoqing_JCP_2023,Chiang_PNAS_2022}, we stress that the procedure and strategy presented is directly applicable to any reference data set, to larger systems, and most importantly, to more complex interfaces.

We evaluated the impact of the different components that are involved in the calculation of the \ac{VSFG} spectra. By comparing the performance of different \ac{XC}, we found that the \ac{PES} is better described by revPBE-D3 and revPBE0-D3 \ac{XC} functionals. However, the \ac{VSFG} spectrum is still 57 cm$^{-1}$ blue-shifted when \ac{NQE}s are properly taken into account. 
These results add to the existing evidence confirming that \ac{XC} functionals constrained by exact functional conditions with suitable dispersion corrections, such as revPBE-D3 and revPBE0-D3, deliver excellent performance in the description of water~\cite{Marsalek_JCPL_2017}.

Since the discrepancies between the theoretical  and the experimental spectra are larger than the training error of the \ac{SA-GPR} models , we attribute the largest source of error to the reference data used to describe the $P$ and $\alpha$ surfaces.
A further study using electronic structure methods beyond \ac{DFT} to obtain more accurate $P$ and $\alpha$ surfaces is urgently needed.

The modelling of the \ac{VSFG} spectra was based on Eq. \ref{eq:TCF} which is derived in the electric dipole approximation \cite{MoritaBook}. We believe that applying the current methodology to a more accurate reference data set would finally resolve existing controversies related to, for example, the relevance of quadrupole contributions in the water bending mode \cite{Kundu_JPCL_2016,Seki_JCPL_2020}. 

Two different methods were considered to describe the time evolution of the nuclei: \ac{MD} and \ac{CMD}. \ac{CMD} is appropriate to describe the D$_2$O/air interface and 
predicts a peak position in good agreement with experiments. 
However, the curvilinear motion of dangling O-H in H$_2$O shows an extremely broad feature \cite{Medders_JACS_2016,Kaliannan_MolPhys_2020}, which calls for the use of other approaches which do not suffer from the curvature problem of \ac{CMD} nor the broadening problem of \ac{TRPMD}. The recent method proposed by Musil \textit{et al.} seems a promising approach to tackle these problems \cite{Musil_JCP_2022}. We note, however, that none of these can describe accurately
the Fermi resonance  contributions~\cite{Xiaoqing_JCP_2023,Althorpe_review_2021,Benson_JCP_2021,Ple_JCP_2021}.

We showed that the off-diagonal components of the $\alpha$ tensor for water are more difficult to learn due to their smaller magnitude when compared to the diagonal ones, which leads to a poorer description of certain $\chi^{(2)}$ matrix elements. Since spherical components mix together  diagonal and off-diagonal elements, it is not possible to train the off-diagonal elements exclusively with current SA-GPR implementations. Future efforts will be directed in this direction, in the development of machine learning models to predict wannier centers \cite{Zhang_PRB_2020}, and in the inclusion of explicit \ac{LR} in the description of the \ac{PES}.

In summary, the work presented here sets a new standard for atomistic simulation of non-linear spectroscopies of condensed phases. While we have focused on the second-order response which is related to 
\ac{VSFG} spectroscopy, the approach presented here paves the way to \textit{ab initio} simulations  of 
2D-VSFG, 2D-IR, 
2D THz-Raman and 2D ThZ-IR-Vis spectroscopies \cite{Perakis_ChemRev_2016,Savolainen_PNAS_2013,Grechko_NatComm_2018}. We expect that applications of the same ideas would bring important atomistic insights into the properties of aqueous interfaces at metallic and biological surfaces \cite{Gonella_NatRev_2021} in solution and under confinement \cite{Kapil_Nature_2022}.

\section*{Materials and Methods}

\subsection*{Molecular Dynamics Simulations}

The water/air interface was modelled using a water slab made of 160 water molecules contained in a 16.63x16.63x44.14 $\AA^3$ simulation box. The simulations were carried out using the i-PI code  \cite{IPI} connected to the LAMMPS package \cite{lammps_2022,Singraber_JCTC_2019}.  using the neural network potential reported in Ref. \cite{Schran_JCP_2020} and \cite{Litman_arxiv_2022}. Unless specified otherwise, the results presented with classical nuclei \ac{MD} were obtained as an average of  20 independent 100 ps trajectories in which the initial configurations were obtained by 100 ps thermalization runs at 300K using the stochastic velocity rescaling thermostat \cite{svr} with a time constant of 200 ps. The partially adiabatic CMD simulations were carried out using the same setup as reported elsewhere \cite{Marsalek_JCPL_2017}.
In all cases, the configurations were saved every 4 fs in the production runs for a posterior calculation of $P$ and $\alpha$ through the  \ac{SA-GPR} or POLY2VS \cite{POLY2VS} models. The \textit{ab initio} trajectories were obtained from a previous work \cite{Ohto_JPCL_2019}.

\subsection*{Symmetry-Adapted GPR Models}

For each dataset, SA-GPR models were built for the polarization $\boldsymbol{P}$ and the $\lambda=0$ and $\lambda=2$ spherical tensor components of the polarizability~\cite{Grisafi2018,Wilkins2019}. We used the hyperparameters of Ref.~\cite{Kapil2020}, which were found to perform well for bulk water.
The supplementary information includes scripts for the training procedure, in which the Cartesian tensor properties of the training set frames are converted to spherical tensors, the $\lambda$-SOAP descriptors of Ref.~\cite{Grisafi2018} are calculated for each local environment in every training configuration and these are combined to give $\lambda$-SOAP kernels, which are used to train separate models for each spherical tensor component.
The supplementary information includes a script that shows in detail how this process works.
In each case, the full data set was split 80\%:20\% with the smaller set used to test the hyperparameters and ensure that the model was not overfit. The final model for each set was trained on the entire set.
The SI also contains each of the models developed and a script that can be used to apply them.
The mixed short-range/long-range models are built using kernels of the form,
\begin{equation}
\boldsymbol{K}^{\lambda}(\mathcal{X},\mathcal{X}') = w_{\rm S} \boldsymbol{K}^{\lambda}_{\text{SR}}(\mathcal{X},\mathcal{X}') + w_{\rm L} \boldsymbol{K}^{\lambda}_{\text{LR}}(\mathcal{X},\mathcal{X}'),
\end{equation}
where $\boldsymbol{K}^{\lambda}(\mathcal{X},\mathcal{X}')$ is the total spherical tensor kernel of order $\lambda$ between two local environments $\mathcal{X}$ and $\mathcal{X}'$, $\boldsymbol{K}^{\lambda}_{\rm SR}$ is the (short-ranged) $\lambda$-SOAP kernel~\cite{Grisafi2018}, $\boldsymbol{K}^{\lambda}_{\rm LR}$ the (long-ranged) $\lambda$-LODE kernel~\cite{Grisafi_JCP_2019} and  $w_{\rm S} + w_{\rm L} = 1$

For the ML-PBE-A model, we used the bulk water data of Ref.~\cite{Grisafi2018}, comprising 1000 frames of 32 molecules, with polarizations and polarizabilities computed using the PBE functional and ultrasoft pseudopotentials.
For these 1000 frames, the polarizations were recomputed at the PBE0 level to produce the ML-PBE0-A set.
These configurations were also used for the ML-POLY-A set, with both the polarization and the polarizability recomputed with the POLY2VS model.
The ML-POLY-B models were produced using a set of water clusters: water $n$-mer configurations (with $n=1,\dots,6$) were taken from the training set used in Ref.~\cite{Inoue2023}: 10,000 clusters were used for each value of $n$, with the POLY2VS model used to calculate the dipole moment $\boldsymbol{\mu}$ and $\boldsymbol{\alpha}$. The predictions of ML-POLY-B give atom-centred dipole moments, which can be summed up to give a prediction of the total polarization of a frame.

The accuracy of SA-GPR models for the polarizability is given by the root mean squared error (RMSE),
\begin{equation}
\text{RMSE}_{\alpha} = \sqrt{\frac{1}{N}\sum_{i=1}^{N} \|\boldsymbol{\alpha}_{\text{calc},i} - \boldsymbol{\alpha}_{\text{pred},i}\|_{2}^{2}},
\end{equation}
where $\|\cdots\|_{2}$ is the Frobenius norm, $N$ the number of members of the training set, $\boldsymbol{\alpha}_{\text{calc},i}$ the calculated polarizability for the $i^{\rm th}$ testing point and $\boldsymbol{\alpha}_{\text{pred},i}$ its predicted polarizability.
For the polarization we use,
\begin{equation}
\text{RMSE}_{P} = \sqrt{\frac{1}{N}\sum_{i=1}^{N} \left\| |\boldsymbol{P}|_{\text{calc},i} - |\boldsymbol{P}|_{\text{pred},i}\right\|_{2}^{2}},
\end{equation}
with $\|\boldsymbol{P}\|_{\text{calc},i}$ the magnitude of the calculated polarization of testing point $i$ and $\|\boldsymbol{P}\|_{\text{pred},i}$ its predicted value.

\begin{acknowledgments}

Y.L. has been partly funded by the Deutsche Forschungsgemeinschaft (DFG, German Research Foundation) project number 467724959. D.M.W. thanks Queen's University Belfast for startup funding. J.L. thanks Simons Foundation Postdoctoral Fellowship. We thank Dr. Venkat Kapil for the useful discussion. We acknowledge the computing resources
from the Swiss National Supercomputing Centre (CSCS) under project ID S1112, S1113.  

\end{acknowledgments} %

\end{document}


\preprint{AIP/123-QED}

\title{Fully First-Principles Surface Spectroscopy with Machine Learning}

\author{Y. Litman}%
\affiliation{Yusuf Hamied Department of Chemistry,  University of Cambridge,  Lensfield Road,  Cambridge,  CB2 1EW,UK}
\affiliation{Max Planck Institute for Polymer Research, Ackermannweg 10, 55128 Mainz, Germany}

\author{Jinggang Lan}
\affiliation{Department of Chemistry, New York University, New York, NY, 10003, USA}
\affiliation{Simons Center for Computational Physical Chemistry at New York University, New York, NY, 10003, USA}
\author{Yuki Nagata}
\affiliation{Max Planck Institute for Polymer Research, Ackermannweg 10, 55128 Mainz, Germany}
\author{David M. Wilkins}
\affiliation{Centre for Quantum Materials and Technology, School of Mathematics and Physics, Queen’s University Belfast, Belfast BT7 1NN, Northern Ireland, United Kingdom}

\maketitle

\section*{Methods}

\subsection*{More details on the calculation of the VSFG spectra}

By introducing Eq. 3 and 4 into Eq. 1, we obtain

\begin{equation}\label{eq:TCF_SI}
\begin{split}
\chi^{(2),R}_{pqr} (\omega_\text{IR}) = i\int_0^\infty dt e^{-i\omega_\text{IR}t}
\sum_
{
\gamma_1,
\gamma_2}\langle g(z_{\gamma_2},z_1,z_2)
 \alpha_{pq}(\mathcal{X}_{\gamma_1},t)P_r(\mathcal{X}_{\gamma_2},0)\rangle,
 \end{split}
\end{equation}
\nid 
where $\alpha_{pq}(\mathcal{X}_{\gamma_1},t)$ is the polarizability of the molecule $\gamma_1$  at time $t$ and $P_r(\mathcal{X}_{\gamma_2},0)$ is the molecular dipole of the molecule $\gamma_2$ at time $0$. The surface normal is assumed to be parallel to the $z$-axis and $z_{\gamma_2}$ corresponds to $z$ coordinate of the $\gamma_2$ water molecule. The results obtained from  the \textit{ab initio} trajectories (Fig. 3 of the main text)  were calculated by considering only the terms $\gamma_1=\gamma_2$ in the sum of Eq. \ref{eq:TCF_SI}. This approximation was necessary due to the limited length of the trajectories.  For all the other results, the sum was evaluated for all pairs of water molecules separated by less than 4~\AA. This cutoff value provides a converged spectra, see for example \cite{Kaliannan_MolPhys_2020}.
Unless specified, we use $z_1=4$~\AA~ and $z_2=2$~\AA~as reported in Ref. \cite{Moberg_JPCB_2018}, and before the calculation of the Fourier transform the correlation function was convoluted with a Hahn window such that it decays to zero after 1.0 ps. Experimental data was processed consistently.

\subsubsection*{Hybrid DFT Calculations}
The hybrid PBE0 calculations were performed using the CP2K program \cite{kuhne2020cp2k}. %
Molecular orbitals of the valence electrons were expanded in the TZV2P basis sets\cite{vandevondele2007gaussian}, while atomic core electrons were described through Goedecker-Teter-Hutter (GTH) pseudopotentials corresponding to the PBE functional \cite{goedecker1996separable,hartwigsen1998relativistic}. Exact exchange integrals were calculated within the auxiliary density matrix method (ADMM) approximation \cite{guidon2010auxiliary}. In addition, the truncated Coulomb operator \cite{truncated} has been applied for the exchange calculations with the cutoff radius approximately equal to half the length of the smallest edge of the simulation cell, together with the Schwarz integral screening with the threshold of $10^{-10}$ a. u. The  cutoff for the auxiliary plane waves was 800 Ry. 

The total electric dipole moment $\mu$ is evaluated using the Berry phase scheme~\cite{king1993theory,resta1994macroscopic}.
With a side length $L$ and considering only the $\Gamma$ point in the Brillouin zone, each component of the electric-dipole moment is then
\begin{equation}
\mu = \frac{e}{2\pi}\mathrm{Im~ln~det}\textbf{S},
\end{equation}
where the matrix \textbf{S} is defined using the Kohn-Sam orbitals $\phi_n$, and
\begin{equation}
S_{n,m}(x) =\int_{L}\phi_{n}^{*}(x) \exp\left[-i\frac{2\pi}{L}x\right]\phi_m(x) dx.
\end{equation}
The polarizatibility $\alpha$ is calculated using the finite difference method, where a periodic electric field is applied in a given direction. 
\begin{equation}
    \alpha = \frac{d\mu}{dE}.
\end{equation}

\section*{Training of SA-GPR models}

\begin{figure}
\centering
\includegraphics[width=\textwidth]{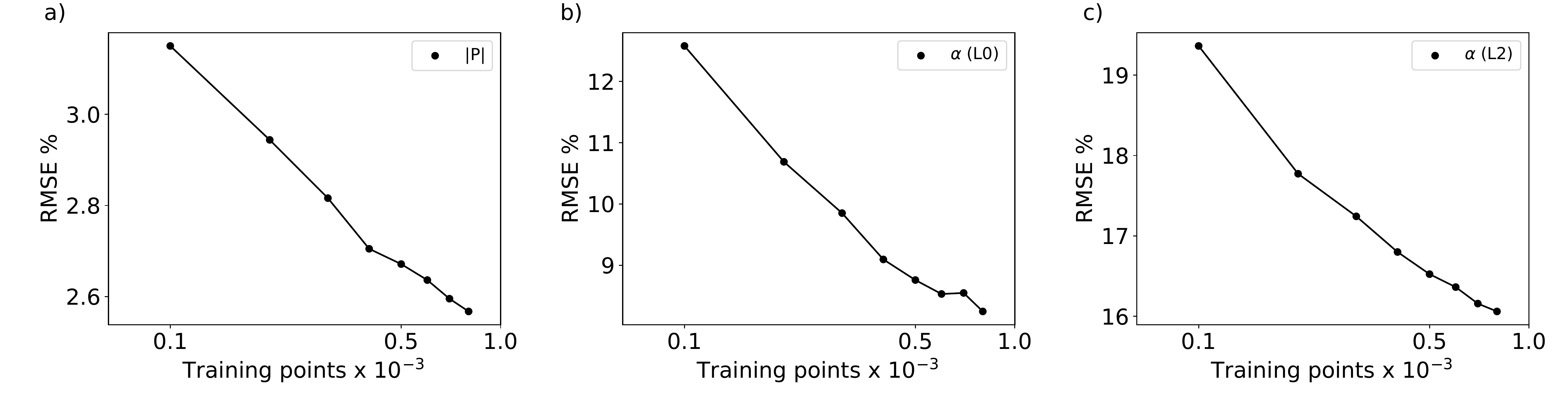}
\caption{ Learning curve of total  polarization (a), $L0$ component of $\alpha$ (b), and $L2$ component of $\alpha$ for 
 ML-POLY-A model}
\label{fig:POLY-A_Learning curves}
\end{figure}

\begin{figure}
\centering
\includegraphics[width=\textwidth]{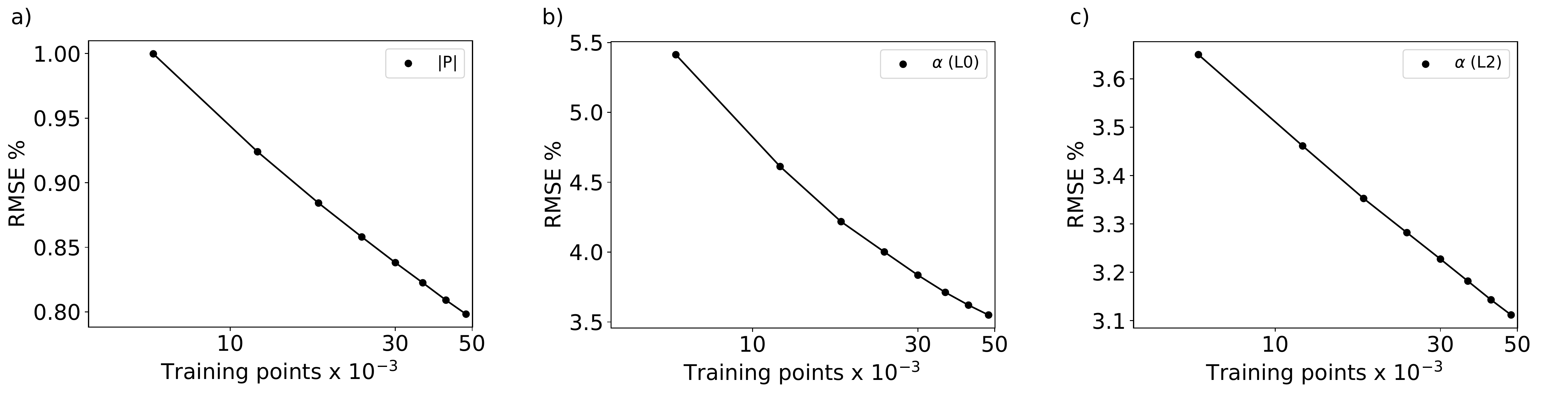}
\caption{ Same as \ref{fig:POLY-A_Learning curves} for  ML-POLY-B model}
\label{fig:POLY-B_Learning curves}
\end{figure}

\begin{figure}
\centering
\includegraphics[width=0.5\textwidth]{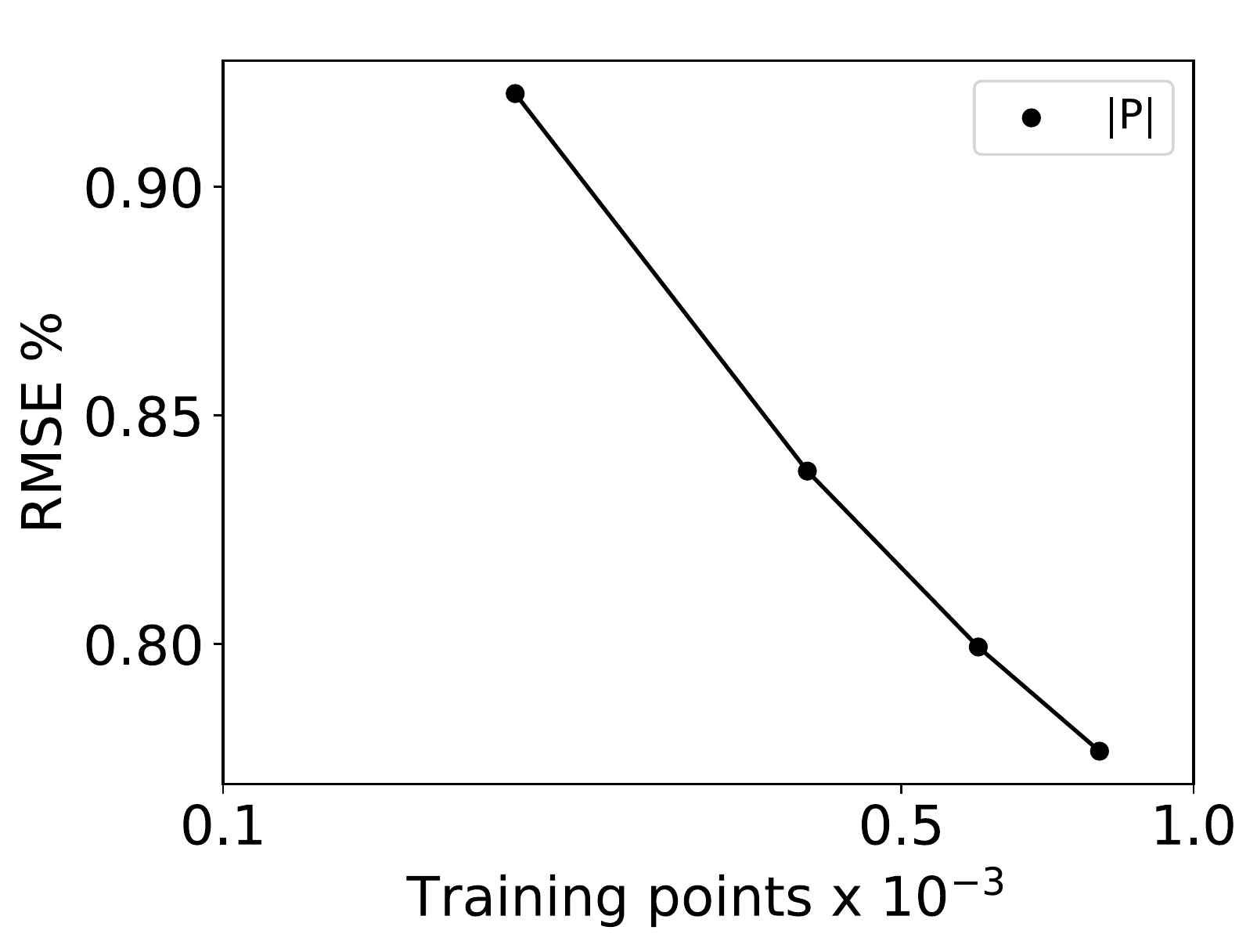}
\caption{ Learning curve of total  polarization fol ML-PBE0-A model.}
\label{fig:PBE0_LC}
\end{figure}

\begin{figure}
\centering
\includegraphics[width=\textwidth]{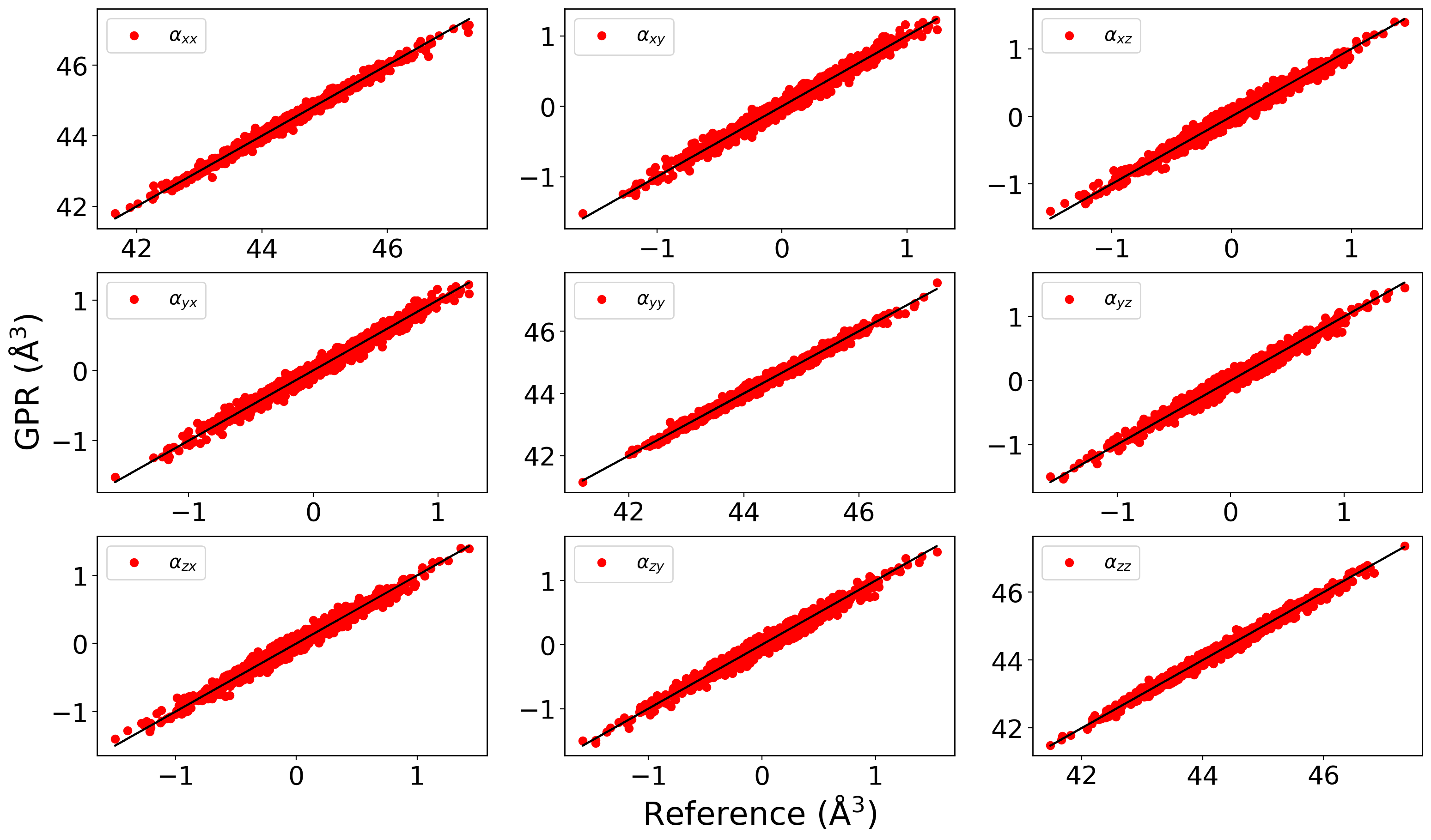}
\caption{Prediction of total $\alpha$ components for ML-POLY-A model.}
\label{fig:POLYA_alpha_total}
\end{figure}

\begin{figure}
\centering
\includegraphics[width=\textwidth]{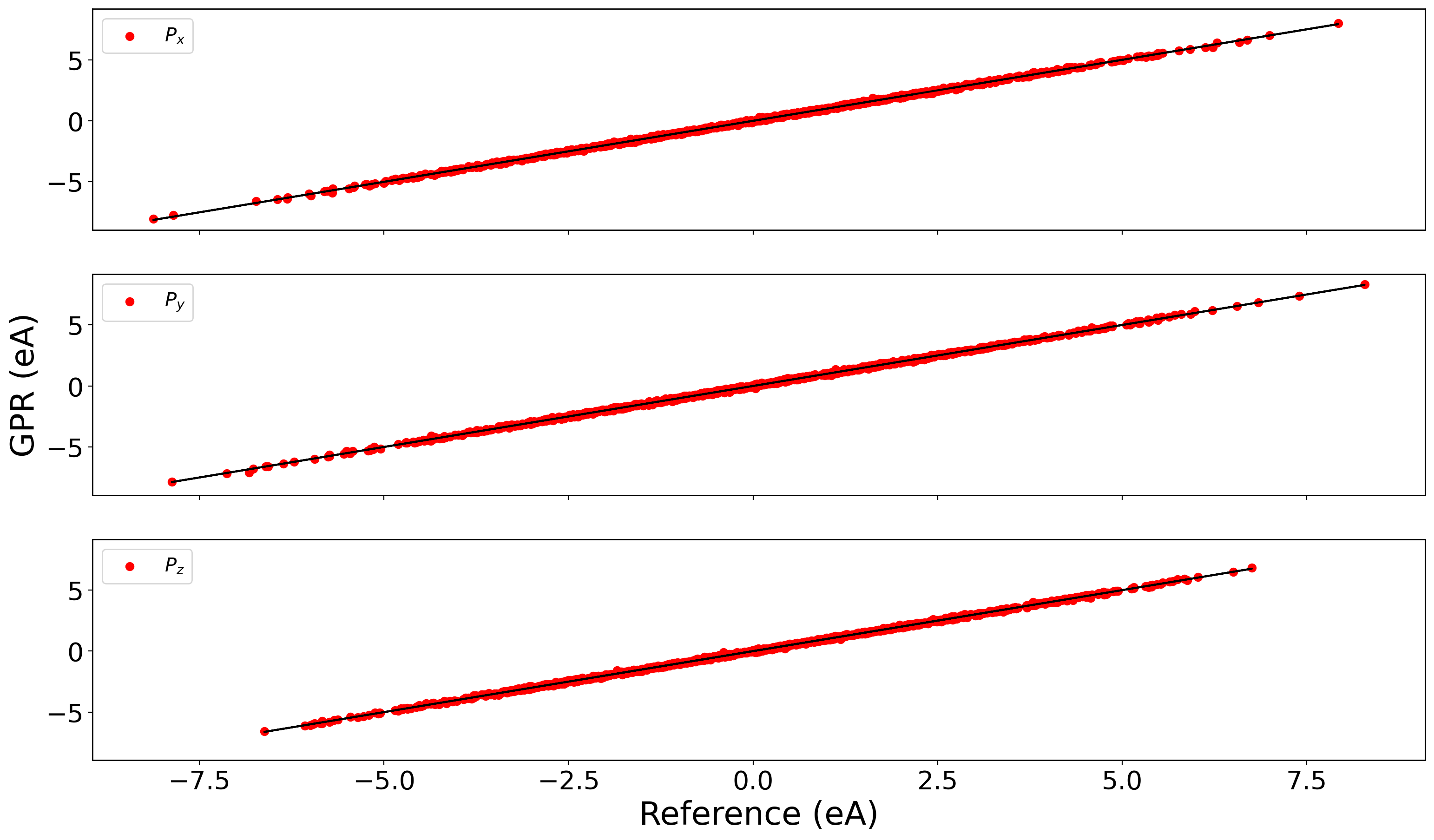}
\caption{Prediction of total $P$ for ML-POLY-A model.}
\label{fig:POLYA_P_total}
\end{figure}

\begin{figure}
\centering
\includegraphics[width=\textwidth]{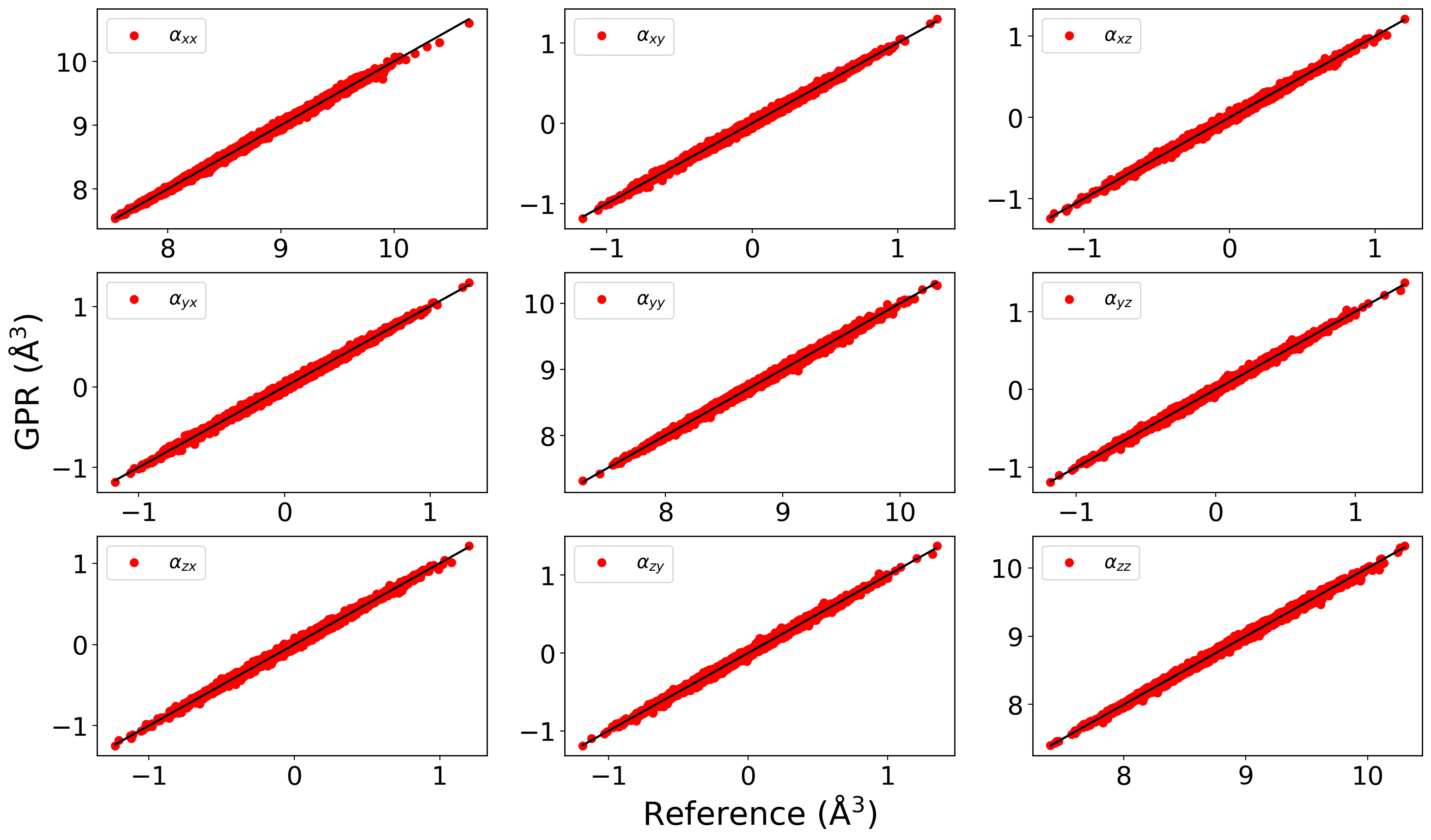}
\caption{Prediction of total $\alpha$ components for ML-POLY-A model.}
\label{fig:POLYB_alpha_total}
\end{figure}

\begin{figure}
\centering
\includegraphics[width=\textwidth]{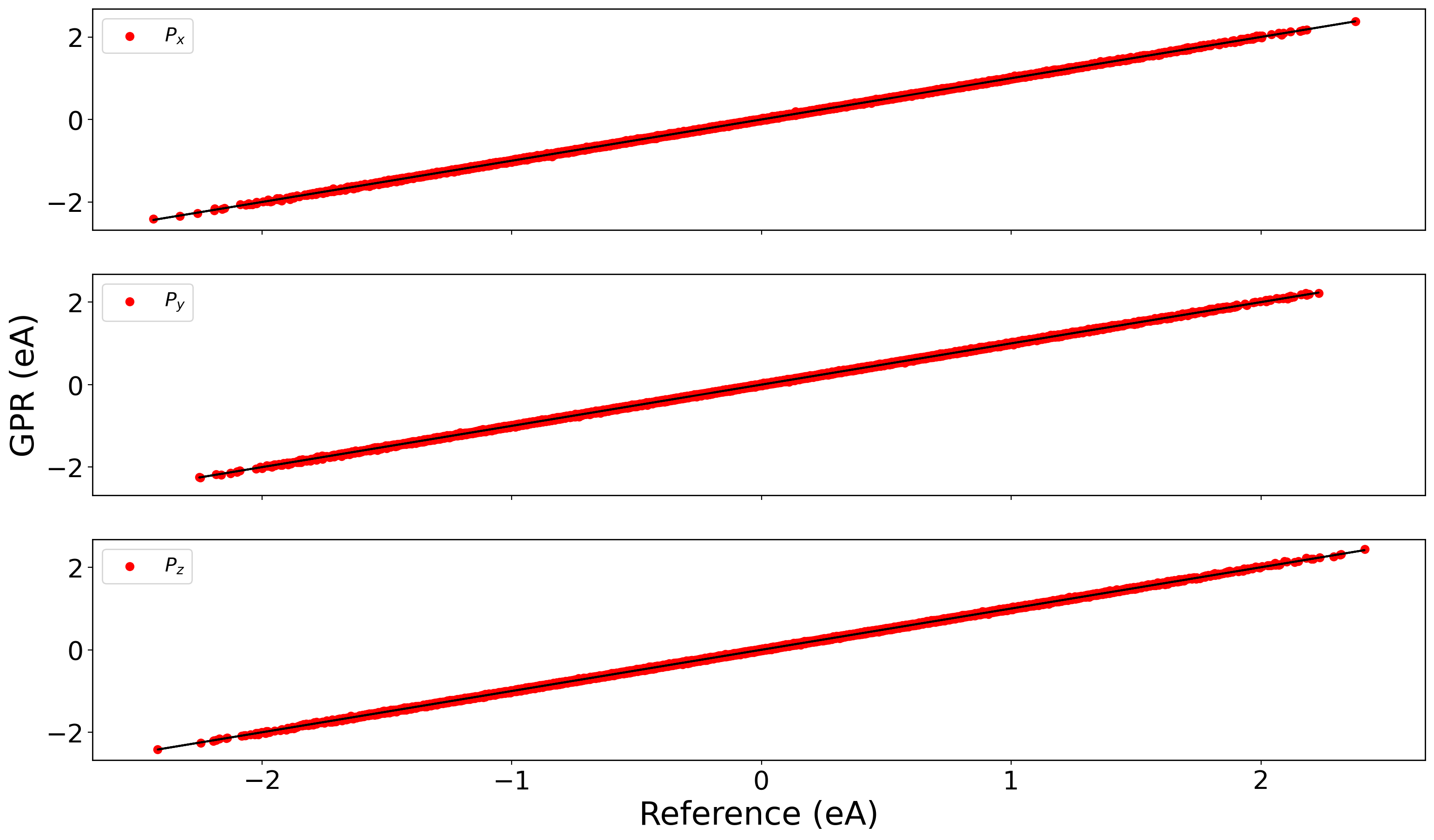}
\caption{Prediction of total $P$ for ML-POLY-B model.}
\label{fig:POLYB_P_total}
\end{figure}

\begin{figure}
\centering
\includegraphics[width=\textwidth]{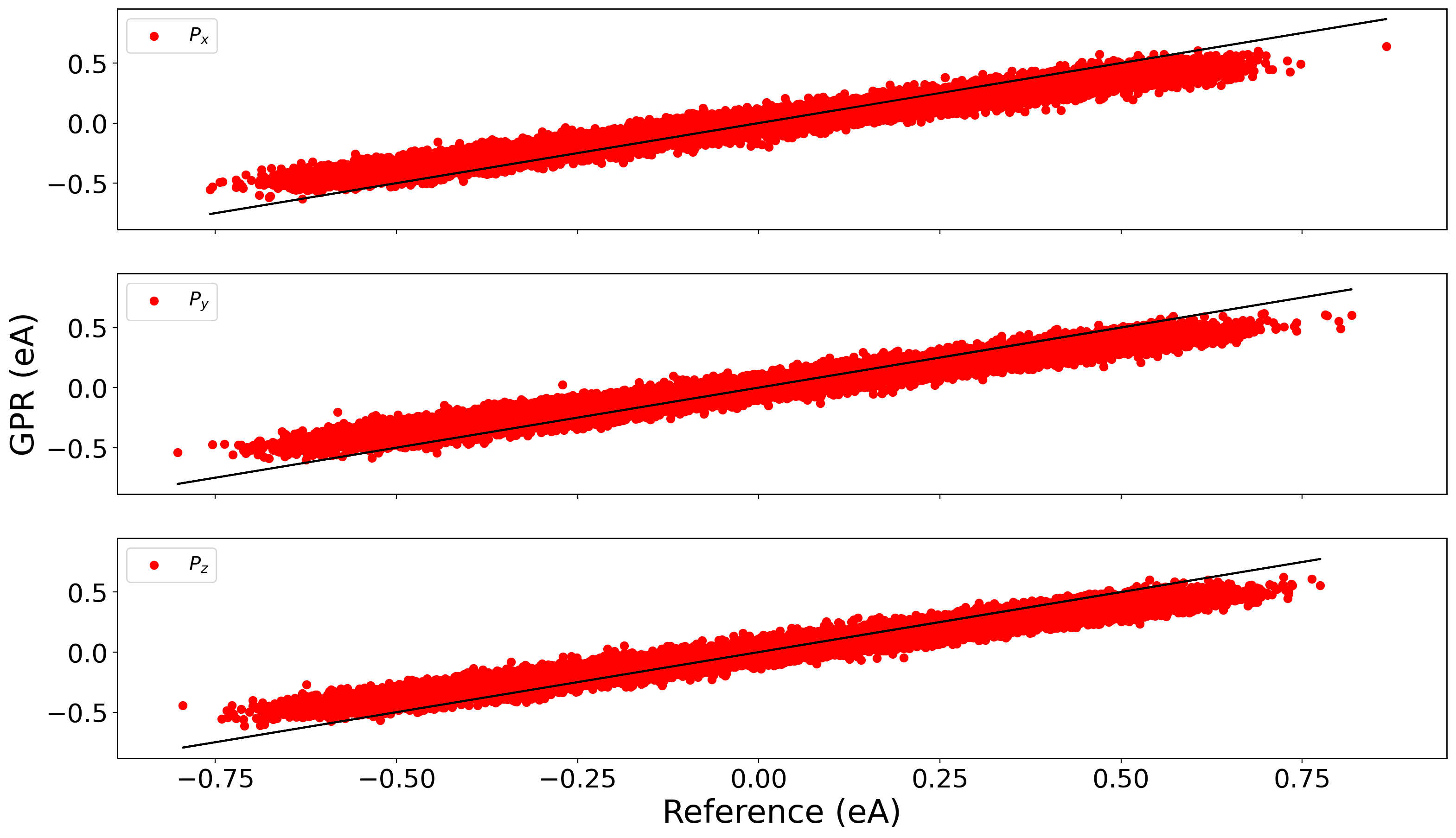}
\caption{Prediction of molecular $P$ of ML-POLY-A model.}\label{fig:POLYA_P_mol}
\end{figure}

\begin{figure}
\centering
\includegraphics[width=\textwidth]{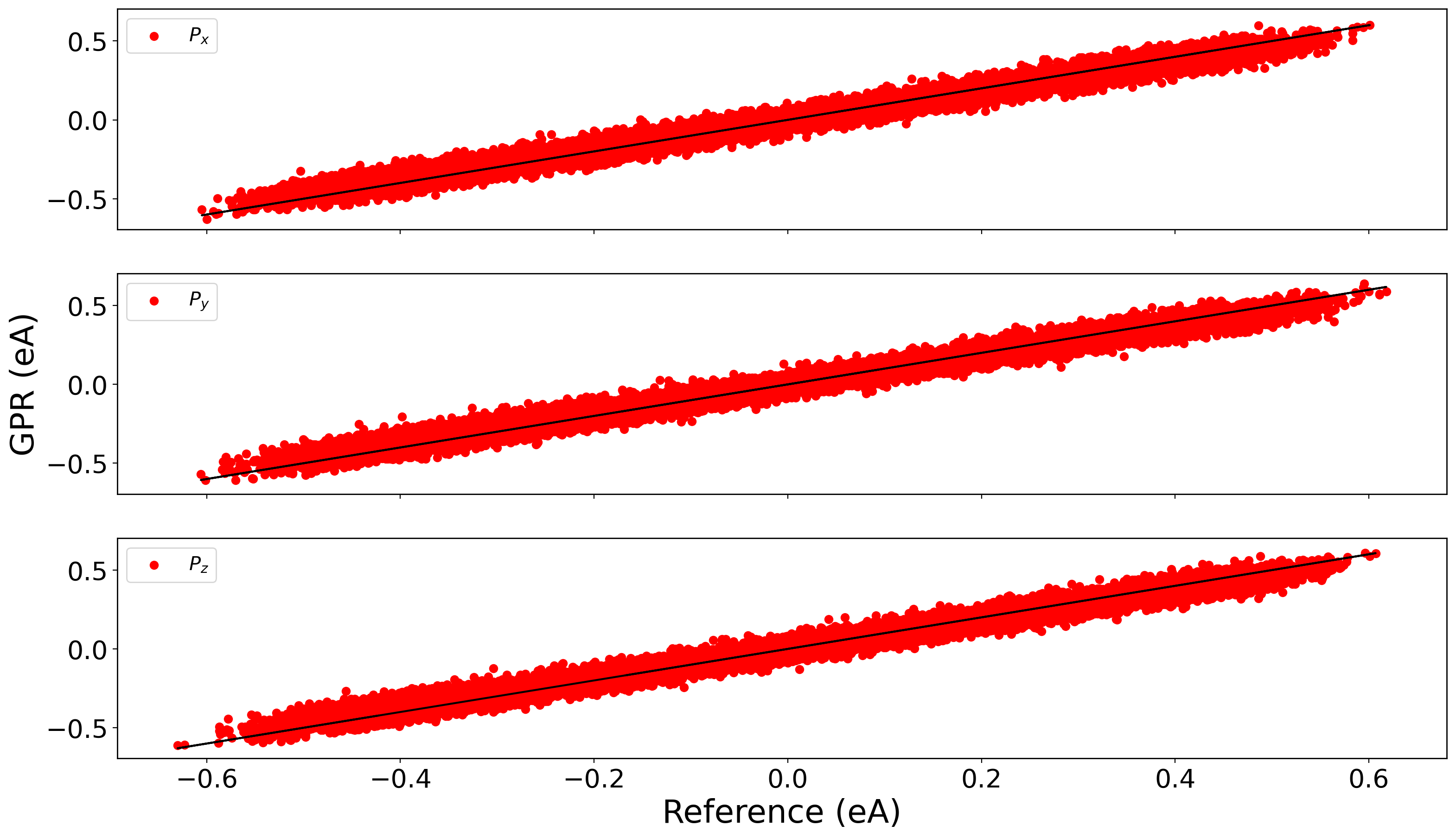}
\caption{Prediction of molecular $P$ of ML-POLY-B model on water hexamers}
\label{fig:POLYB_P_mol}
\end{figure}

The learning curves for polarization ($P$) and polarizability ($\alpha$) for ML-POLY-A and ML-POLY-B models are presented in Fig. 
\ref{fig:POLY-A_Learning curves} and 
\ref{fig:POLY-B_Learning curves}, respectively.
The learning curves for ML-PBE-A can be found elsewhere \cite{Kapil2020} and $P$ learning curve for ML-PBE0-A is presented in Fig. 
\ref{fig:PBE0_LC}.
 In Fig. 
\ref{fig:POLYA_alpha_total} (\ref{fig:POLYB_alpha_total}) and 
\ref{fig:POLYA_P_total} (\ref{fig:POLYB_P_total}), we present the per-component correlation plots  for total $\alpha$ and $P$ predictions for the ML-POLY-A (ML-POLY-B) model.
All these figures demonstrate that \ac{SA-GPR} can handle cluster and bulk structures equally well.
 Since the POLY2VS force-field uses a molecular decomposition of the $P$, and \ac{SA-GPR} utilizes an atomic decomposition underneath, it is possible to analyze how well the molecular quantities are being represented.
In Fig. \ref{fig:POLYA_P_mol} and \ref{fig:POLYB_P_mol}, we present the corresponding plots for the molecular predictions.
The errors obtained  for molecular quantities are much larger than the ones obtained for the total  quantities. This fact is not surprising since our models are trained on real observables (i.e. total $P$ and total $\alpha$) and we are not using any physical constraints on our models beyond the symmetry considerations. 

\begin{figure}
\centering
\includegraphics[width=\textwidth]{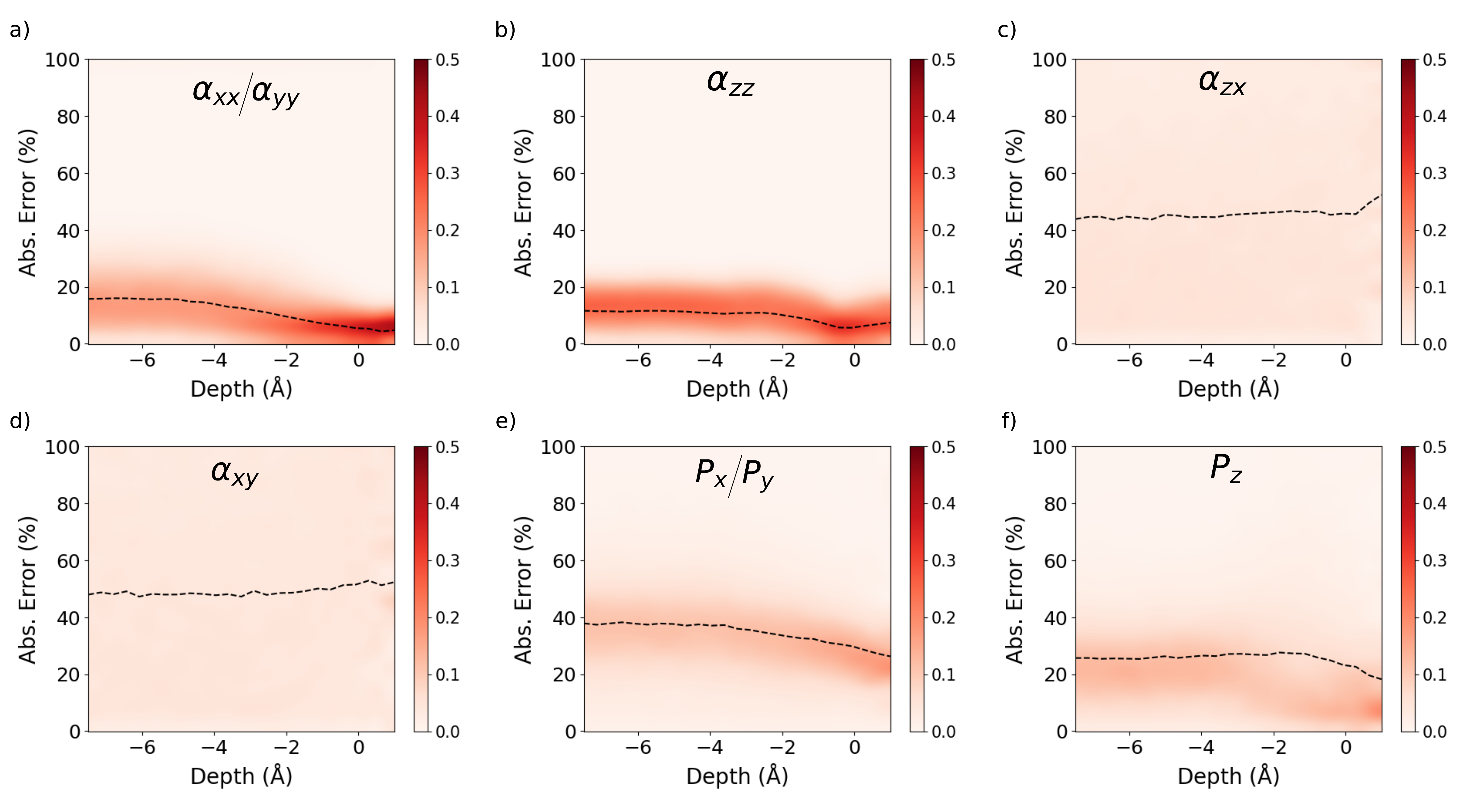}
\caption{Absolute error histograms for molecular  $\alpha$ and $P$ predictions as a function of distance to the interface using the ML-POLY-A model. Depth distances were computed with respect to the instantaneous interface \cite{Willard_JPCB_2010}. 
Black dashed lines correspond to average values.}
\label{fig:Depth_POLY_A}
\end{figure}

\begin{figure}
\centering
\includegraphics[width=\textwidth]{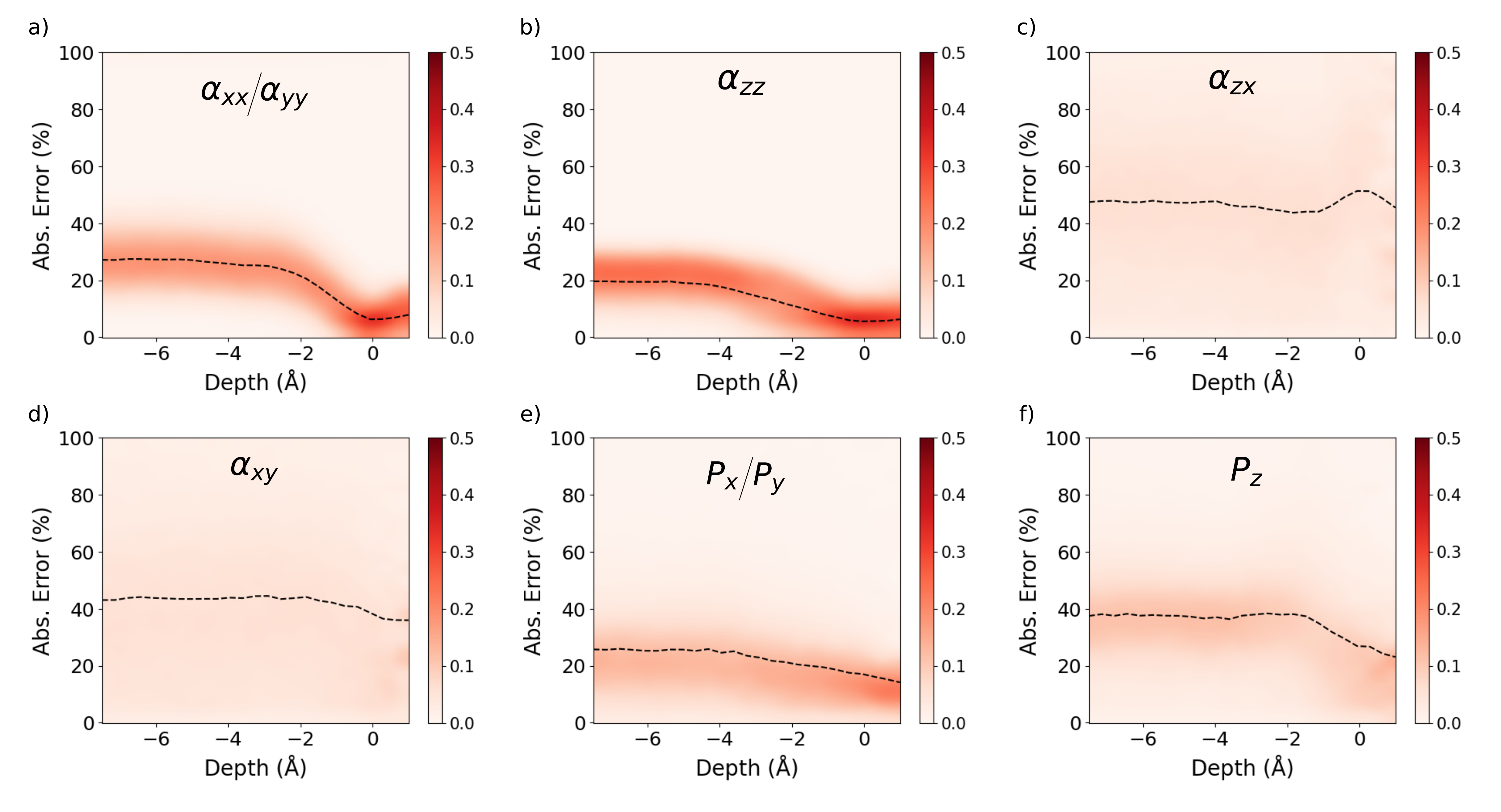}
\caption{Same as \ref{fig:Depth_POLY_A} for ML-POLY-B model. }
\label{fig:Depth_POLY_B}
\end{figure}

In Fig. \ref{fig:Depth_POLY_A} and Fig. 
\ref{fig:Depth_POLY_B}, we present the absolute percentage errors of the molecular predictions for ML-POLY-A and ML-POLY-B models, respectively, as a  function of distance from the interface in the same setup used to compute the \ac{VSFG} spectra, i.e. water slab structures. Note that these structures are not included in the training sets. Both models show similar performance, with ML-POLY-B performing slightly better, even though ML-POLY-A was trained on bulk structures exclusively, and surprisingly the predictions are more accurate in the vicinity of the interface.  
While the error distributions for the diagonal components of $\alpha$ are relatively narrow around the mean value, the off-diagonal elements show a more uniform distribution with errors up to 100\%. We believe that the larger errors in the latter case are due to the smaller relative values of the off-diagonal elements with respect to the diagonal ones. 
 The polarization components show a better performance than the off-diagonal elements but worse than the diagonal elements of $\alpha$. Note that, as it is shown in the main text (and in Fig. \ref{fig:Cross-POLY} below),  all the $\chi^{(2)}$ component are predicted satisfactorily except $\chi^{(2)}_\text{zxx}$ which utilizes one of the off-diagonal components of $\alpha$. 

\section*{Additional Figures}

\begin{figure}
\centering
\includegraphics[width=\textwidth]{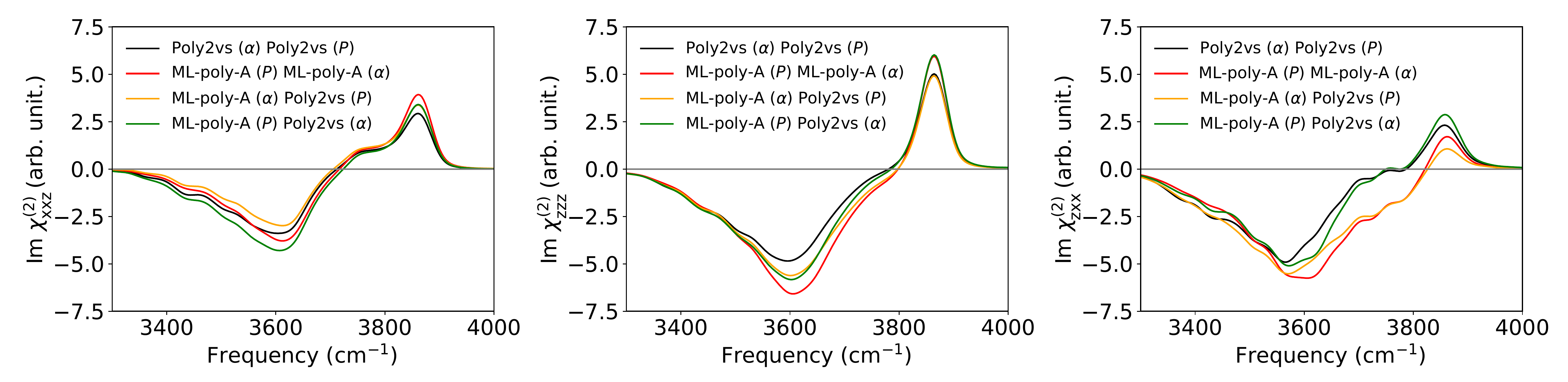}
\caption{
Imaginary part of non-zero and independent $\chi^{(2)}$ components of the water/air interface using different combinations of 
 POLY2VS and ML-POLY-A  $P$ and $\alpha$ surfaces for a slab geometry made of   160 water molecules.  This test was performed for a shorter trajectory, thus the increased noise in the signal with respect other figures. 
}\label{fig:Cross-POLY}\end{figure}

\begin{figure}
\centering
\includegraphics[width=0.5\textwidth]{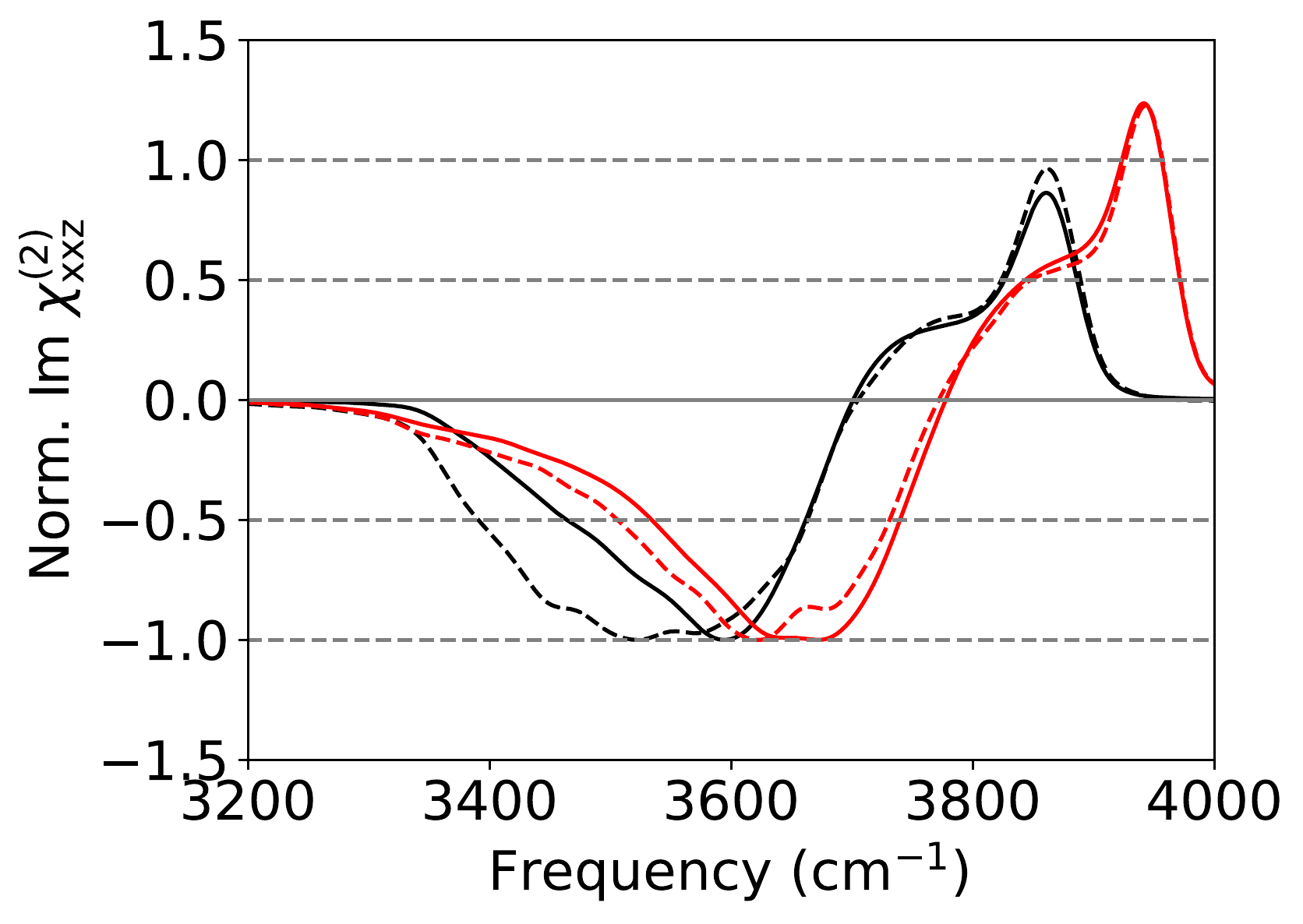}
\caption{Imaginary part of $\chi^{(2)}_\text{xxz}$ of the water/air interface using the  POLY2VS $P$ and $\alpha$ surfaces combined with POLY2VS potential energy surface (PES) (black) and  HDNNP trained on revPBE0-D3 data PES (red). Solid and dashed lines correspond to calculation  1 \AA~ and 4 \AA~ cutoff values, respectively. Data has been normalized at the H-bonded band to ease comparison.}\label{fig:NN_with_POLY2VS}\end{figure}

\begin{figure}
\centering
\includegraphics[width=0.7\textwidth]{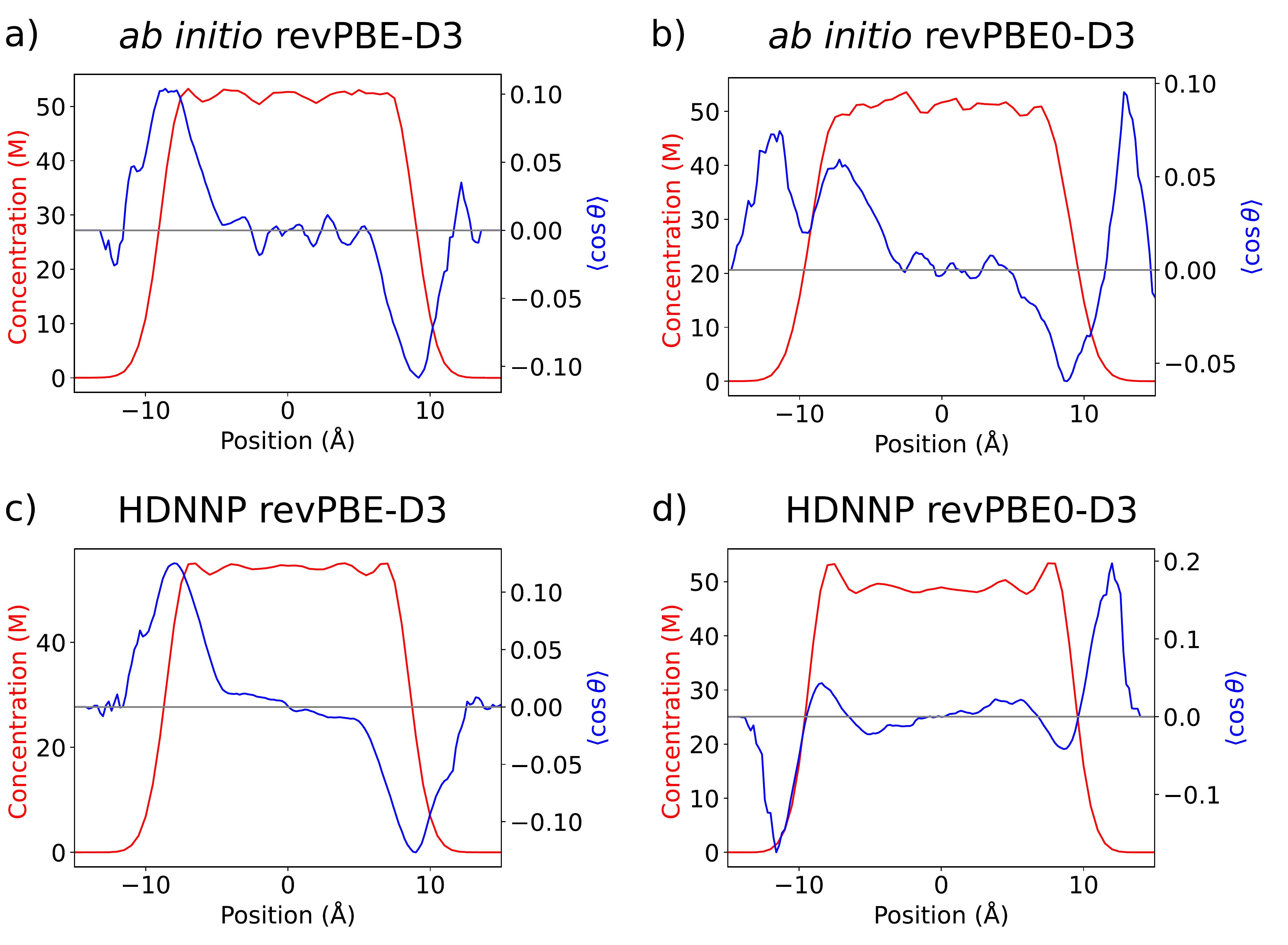}
\caption{Water concentration and $\cos(\theta)$ profiles along the direction orthogonal to the slab surface. The center of the slab is set 0 \AA. $\theta$ corresponds to the angle between the water bisector and the direction orthogonal to the water/air interface.
}\label{fig:angles}\end{figure}

\begin{figure}
\centering
\includegraphics[width=\textwidth]{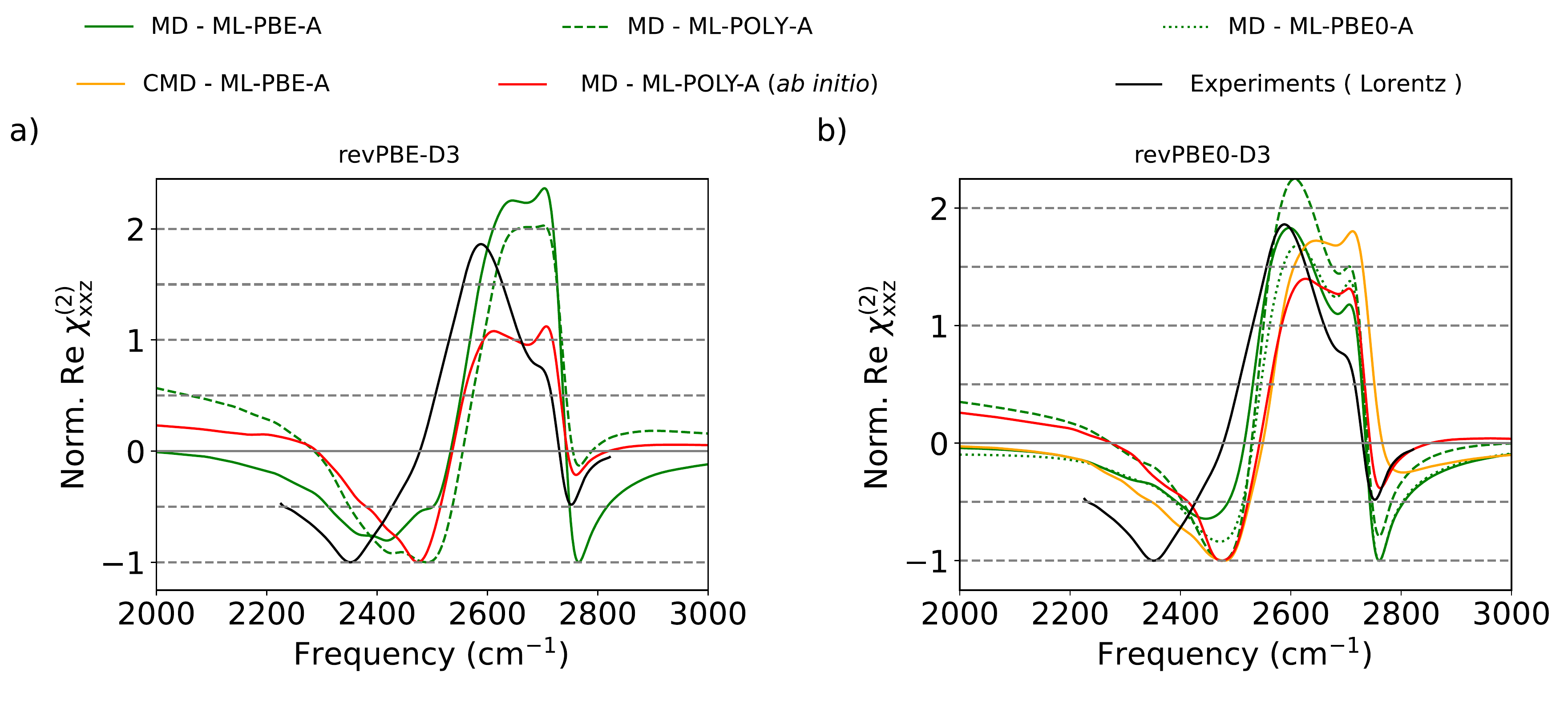}
\caption{Normalized Real part of $\chi_{xxz}$  spectra of the D$_2$O/air interface at 300K.  
Simulated spectra using revPBE-D3 (left) and revPBE0-D3 (right) \ac{XC} functionals.
Classical \ac{MD} simulations using \ac{HDNNP}s are presented  
by solid (ML-PBE-A), 
, dashed (ML-POLY-A), and dotted (ML-PBE0-A) green lines.
\ac{CMD} simulations are depicted with solid orange lines while
results based on direct \textit{ab initio} trajectories are depicted with red lines.
Experimental spectra are depicted with solid black lines and horizontal gray lines have been added to guide the eye. 
Spectra have been rigidly shifted to match experimental spectra.
 Experimental spectra are corrected by the appropriate Fresnel factors, assuming the Lorentz model (black dotted lines) for the interfacial dielectric constant \cite{Xiaoqing_JCP_2023}. To allow a comparative analysis,  we present the spectra, normalized such that the minimum intensity has an intensity of minus one.}
\label{fig:Real_part}
\end{figure}

\begin{figure}
\centering
\includegraphics[width=0.9\textwidth]{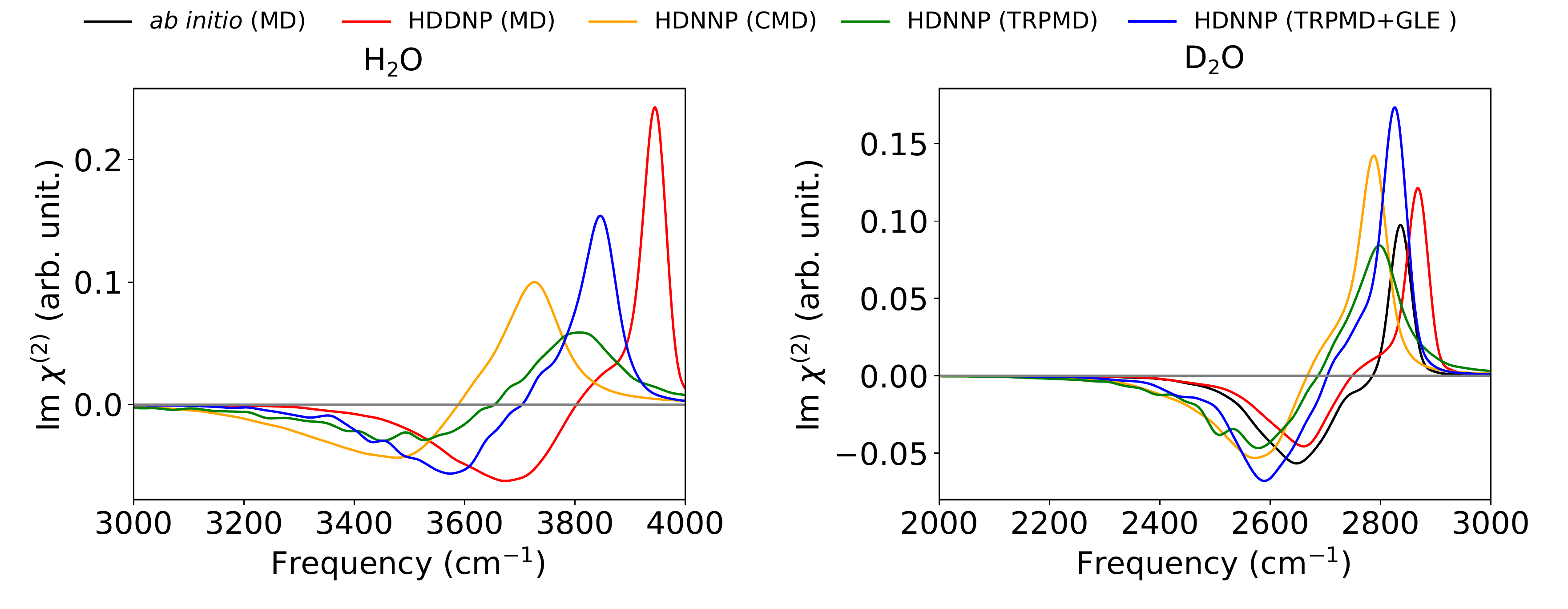}
\caption{Imaginary part $\chi^{(2)}$ of the water/air interface calculated using the ssVVCF approximations \cite{ssVVCF} and different molecular dynamics (MD, red line),
centroid molecular dynamics (CMD, orange line),
thermostated ring polymer molecular dynamics (TRPMD, orange line),
and thermostated ring polymer molecular dynamics tuned with a generalized Langevin thermostat (TRPMD+GLE, blue line). All simulations were performed at 300K employing the HDNNP trained on revPBE0-D3 data. }\label{fig:NQEs}\end{figure}

In Fig. \ref{fig:Cross-POLY}, we present calculations for the imaginary part of $\chi^{(2)}$ using GPR and POLY2VS $\alpha$-$P$ surfaces using trajectories obtained with POLY2VS potential energy surface (PES). In Fig. \ref{fig:NN_with_POLY2VS}, we compare the predicted spectra using different PES but the same  POLY2VS $\alpha$-$P$ surfaces using different cutoff values (see Eq. \ref{eq:TCF_SI}).
In Fig. \ref{fig:angles}, we compare the density and orientation profiles obtained with \textit{ab initio} and neural-network PES.
In Fig. \ref{fig:Real_part} we present the real part of $\chi^{(2)}$ corresponding to the simulations shown in Fig. 3 in the main text. Finally, in Fig. \ref{fig:NQEs} we present imaginary part of $\chi^{(2)}$ obtained using three different methods to propagate the nuclei: classical nuclei molecular dynamics (MD), 
thermostated ring polymer molecular dynamics (TRPMD) \cite{TRPMD,TRPMD_GLE} and  centroid molecular
dynamics (CMD) \cite{CMD_I,CMD_II}.
\vfill
\pagebreak

 \clearpage